\documentclass[aps,prd,twocolumn,amsmath,superscriptaddress,amssymb,showpacs,floatfix,nofootinbib,longbibliography]{revtex4-1}

\usepackage{graphicx}
\usepackage{dcolumn}
\usepackage{xcolor}
\usepackage{bm}
\usepackage{natbib}
\usepackage{calligra}
\usepackage[T1]{fontenc}
\usepackage{egothic}
\usepackage[T1]{fontenc}
\newfont{\rsfsten}{rsfs10 scaled 1200}
\newfont{\rsfsseven}{rsfs10 scaled 1200}
\newfont{\rsfsfive}{rsfs10 scaled 1200}
\usepackage{epsfig}
\usepackage{units}
\usepackage[utf8]{inputenc}

\newcommand{\aap}{{Astron.~Astrophys.}}
\newcommand{\apjl}{{Astrophys.~J.~Lett.}}

\newcommand{\mnras}{{Mon.~Not.~R.~Astron.~Soc.}}

\newcommand{\be}{\begin{equation}}
\newcommand{\ee}{\end{equation}}
\newcommand{\bea}{\begin{eqnarray}}
\newcommand{\eea}{\end{eqnarray}}

\def\lsim{\mathrel{\raise.3ex\hbox{$<$\kern-.75em\lower1ex\hbox{$\sim$}}}}
\def\gsim{\mathrel{\raise.3ex\hbox{$>$\kern-.75em\lower1ex\hbox{$\sim$}}}}

\begin{document}

\title{Studying the Milky Way Pulsar Population with Cosmic-Ray Leptons}

\author{Ilias Cholis}
\author{Tanvi Karwal}
\author{Marc Kamionkowski}
\affiliation{Department of Physics and Astronomy, The Johns Hopkins University, Baltimore, Maryland, 21218, USA}

\date{\today}

\begin{abstract}
Recent measurements of cosmic-ray electron and positron spectra
at energies from a GeV to 5 TeV, as well as radio, X-ray and a wide 
range of gamma-ray observations of pulsar-wind nebulae, indicate 
that pulsars are significant sources of high-energy cosmic-ray 
electrons and positrons.  Here we calculate the local cosmic-ray 
$e^\pm$ energy spectra from pulsars taking into account 
models for (a) the distribution of the pulsars spin-down properties; 
(b) the cosmic-ray source spectra; and (c) the physics of
cosmic-ray propagation.  We then use the measured cosmic-ray
fluxes from \textit{AMS-02}, \textit{CALET} and 
\textit{DAMPE} to constrain the space of pulsar and
cosmic-ray-propagation models and in particular, local
cosmic-ray diffusion and energy losses, the pulsars' energy-loss
time-dependence, and the injected $e^{\pm}$ spectra.  We find
that the lower estimates for the local $e^{\pm}$ energy losses
are inconsistent with the data.  We also find that pulsar
braking indexes of 2.5 or less for sources with ages more than
10 kyr are strongly disfavored.  Moreover, the cosmic-ray data are
consistent with a wide range of assumptions on the $e^{\pm}$ 
injection spectral properties and on the distribution of initial
spin-down powers. Above a TeV in energy, we find that pulsars 
can easily explain the observed change in the $e^{+} + e^{-}$ 
spectral slope. These conclusions are valid as
long as pulsars contribute $\gtrsim10\%$ of the observed
cosmic-ray $e^\pm$ at energies $\gtrsim100$~GeV.
\end{abstract}


\maketitle

\section{Introduction}
\label{sec:introduction}

Observations of of electromagnetic radiation from pulsars and
their surrounding environment, including the pulsar wind nebulae
(PWNe), from radio wavelengths to $\gamma$-rays
\cite{Kuiper:2001ev, Kuiper:2003an, Thompson:2003tj,
Gaensler:2006ua} suggest that pulsars are a significant source
of high-energy cosmic-ray electrons and positrons.  In particular, HAWC
\cite{Abeysekara:2017hyn, Abeysekara:2017old} and Milagro
\cite{Abdo:2009ku} both recently observed $\gamma$-ray halos at
energies of 10 TeV and above around Geminga and Monogem, two
nearby pulsars. These observations are well accounted for by the
escape of cosmic-ray $e^{\pm}$ from the relevant PWNe
which then produce the observed gamma-rays via inverse-Compton
scattering (ICS) of background light within a volume of $\sim
10^{3}$ pc$^{3}$ around the pulsar \cite{Hooper:2017gtd,
Abeysekara:2017old}.  Follow-up observations will soon address
remaining uncertainties in the diffusion and energy losses of
these leptons in the interstellar medium (ISM) and the possible
effects of convective winds around Geminga and Monogem.
Still, current data already indicate that pulsars and PWNe can
accelerate significant fluxes of $e^{\pm}$ with potential
implications for future pulsar searches \cite{Linden:2017vvb}. 

Cosmic-ray electrons are also thought to be shock-accelerated to
energies between a keV and $\sim100$ TeV in supernova remnants.
At low energies, cosmic-ray electrons and positrons may also
be produced from inelastic collisions of cosmic-ray nuclei with
nuclei in the ISM.  These are commonly
known  as secondary electrons and positrons, and numerical codes 
calculating their spectra have been developed e.g., in Refs.~\cite{Moskalenko:2001ya,  
Kachelriess:2015wpa, GALPROPSite, Evoli:2008dv, DRAGONweb}. 
 
There is roughly one pulsar born in the Galaxy per century
\cite{1999MNRAS.302..693D, Vranesevic:2003tp,
FaucherGiguere:2005ny,Lorimer:2006qs, Keane:2008jj}. Electrons
and positrons suffer from energy losses due to synchrotron
radiation and ICS off the cosmic microwave background (CMB) and
infrared/optical starlight as they diffusively propagate through the ISM.
The interplay of diffusion and energy losses gives a rough maximum energy 
$E_{\rm max} \sim 100\, {\rm GeV} (R/2\, {\rm kpc})^{-2}$
\cite{Cholis:2017ccs} for $e^{\pm}$ that survive at a distance
$R$ from their source.  Thus, fewer sources
can contribute to the $e^\pm$ flux observed at higher energies
at any given location.  A rate of one pulsar per century 
suggests that only a few dozen pulsars contribute to the $e^\pm$
flux above 500 GeV.  The discreteness of the source population
can result in spectral features in the $e^{\pm}$ energy spectra
\cite{Malyshev:2009tw,Grasso:2009ma} that  might be sought,
e.g., with a fluctuation analysis of the energy spectra
\cite{Cholis:2017ccs}.

The aim of this paper is to use existing measurements of the
$e^{\pm}$ energy spectra to constrain the properties of the
pulsar population within a few kpc from the Earth.  We do so
by simulating a large number of realizations of pulsar
distributions for an array of models of the astrophysical conditions 
impacting the cosmic-ray spectra from pulsars.  We first simulate the 
spatial distribution of pulsars. Then for each simulation, we calculate 
the local CR $e^\pm$ spectrum for an array of different assumptions on the 
injected $e^\pm$ spectra and cosmic-ray propagation conditions.  By 
requiring the local cosmic-ray $e^\pm$ energy spectra to agree with
measurements, we exclude over three quarters of the models and
find several conclusions that can be drawn even after
marginalizing over the model uncertainties.  These conclusions
include that braking indexes of 2.5 or less, that have been observed
for some very young pulsars, are excluded by CR data that rely on 
the characteristics of sources older than 10 kyr. Furthermore, we show
that if the local ISM conditions result in low energy losses, then pulsars 
can not explain the data. If such conditions are avoided, pulsars can 
explain the CR data with the positron fraction above 300 GeV being 
either flat, increasing or decreasing with energy. Additionally, a total
$e^{+} + e^{-}$ spectrum with a softer slope is a typical expectation 
of pulsar sources if very young sources such as Vela are still subdominant
contributors in the local $e^{\pm}$.

This paper is organized as follows:  In Section~\ref{sec:method}
we describe the simulations, enumerate the assumptions
made, and clarify the astrophysical uncertainties involved in
the simulations.  In Section~\ref{sec:results} we present the
simulations that are allowed by the data.  We discuss first our
fits to the \textit{AMS-02} positron fraction
($e^{+}/(e^{+}+e^{-})$) measurement. Then we show the impact of
adding into our analysis $e^{+}+e^{-}$ fluxes from
\textit{CALET} and \textit{DAMPE}. We conclude and discuss
future directions in Section~\ref{sec:conclusion}.

\section{Method}
\label{sec:method}

\subsection{Cosmic Ray Data}

We use the published \textit{AMS-02} data from
Ref.~\cite{Accardo:2014lma} collected over a period of 2.5
years. We ignore the measurement associated with energies below
5 GeV since at these energies the spectra are strongly affected
by the solar wind and because pulsars contribute marginally.  We
focus instead on the measurement above 5 GeV and up to 500
GeV.  In addition, we consider the impact of a turnover in the
positron fraction above 500 GeV, as suggested recently
\cite{AMSLaPalma}.  Since this is not a published measurement,
we indicate the impact if the result stands, but also provide
results without it. In addition, we use the 1.5 years of
spectral measurements of the combined $e^{+} + e^{-}$ CR flux,
from 25 GeV up to 5 TeV, by \textit{DAMPE} \cite{TheDAMPE:2017dtc, 
Ambrosi:2017wek} as well as the same spectral measurements by \textit{CALET} 
\cite {2015JPhCS.632a2023A, Adriani:2017efm, Adriani:2018ktz} at
energies between 10 GeV and 5 TeV, taken over an extension of
two years. We note that the measurements by \textit{DAMPE} and
\textit{CALET} are in statistical tension with each other and
thus avoid fiting to both data sets. Instead, we only check for
consistency between the \textit{AMS-02} positron fraction and each
of these electron- plus-positron spectra separately.  

\subsection{Studying the Pulsar Population}

\begin{figure}
\hspace{0.0cm}
\includegraphics[width=3.7in,angle=0]{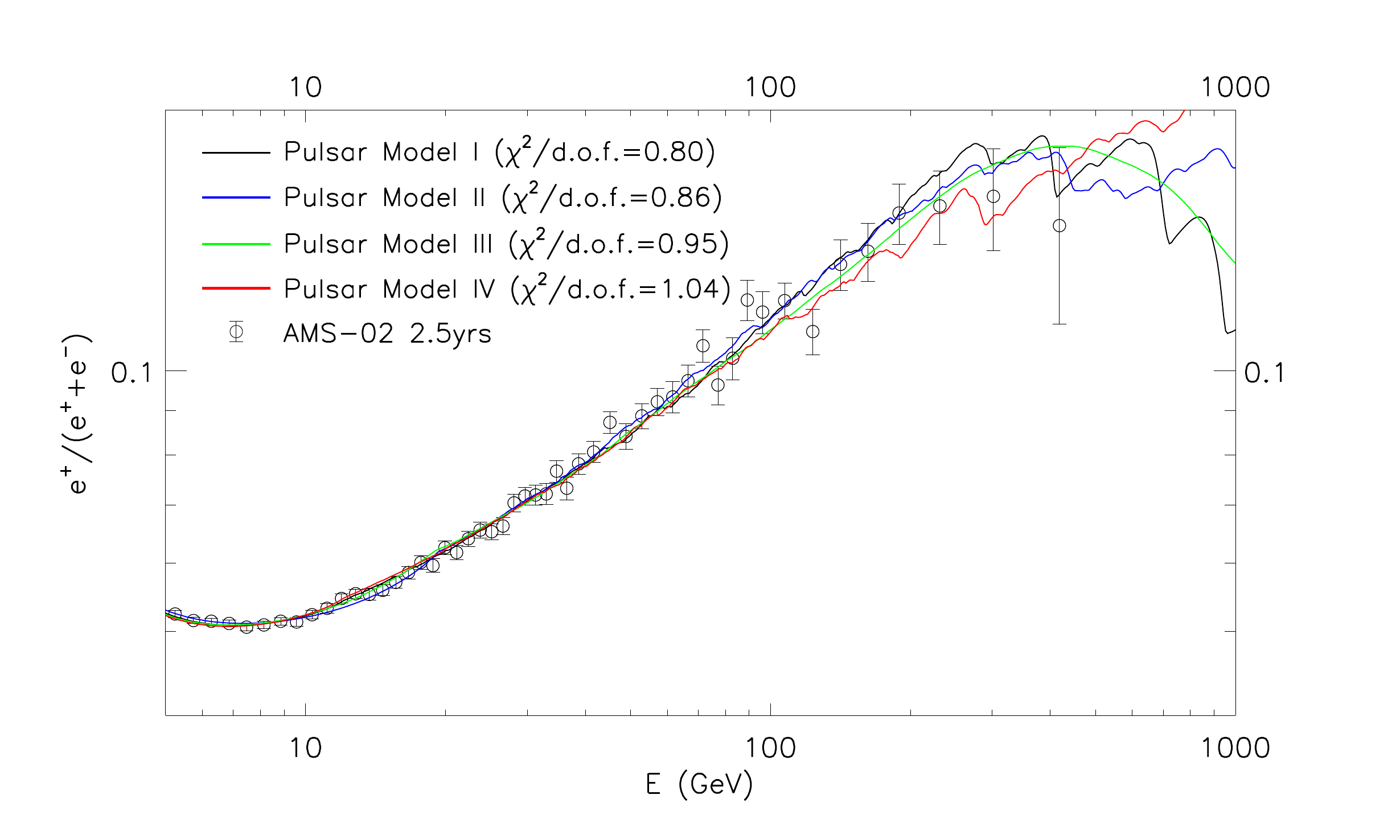}
\vspace{-0.4cm}
\caption{Four different pulsar astrophysical realizations. The
     predicted positron-fraction spectrum, which is observed to
     increase from 7 to $\sim$300 GeV, can either drop at higher
     energies (model I, in black line), flatten out (model II,
     in blue line), or keep rising up to a TeV (model IV, in red
     line). Also depending on the exact ISM assumptions (see
     text for details and~Fig.~\ref{fig:AstroRealiz}), it can be
     smooth (green line) or have features associated with
     individual pulsars whose presence can potentially be
     detected \cite{Cholis:2017ccs}.} 
\label{fig:FourPulsarModels}
\end{figure}

The contribution of local pulsars to the measured CR spectra is influenced by 
uncertainties. We model these uncertainties by producing 
astrophysical realizations spanning the relevant multi-dimensional
parameter space in a discrete manner. We  
call these \textit{unique} points on this space \textit{pulsar
astrophysical realizations} or just \textit{simulations}.  An
example of four such pulsar astrophysical realizations is shown
in Fig.~\ref{fig:FourPulsarModels}. The current data show
that the positron fraction rises monotonically from 7 GeV 
to $\sim$300 GeV.  We show several simulations that fit the
data over the range that it is measured.  These fitted
simulations show that the spectrum can either continue to rise
to a value $\gsim 0.20$ at an energy of a TeV or flatten to a
value of $\simeq 0.16$, or even drop to $\simeq 0.1$ at 
that energy. $\textit{AMS-02}$ will have the sensitivity to
eventually observe such values of the positron
fraction.  Moreover, different astrophysical realizations can
predict a pulsar spectrum that is inherently smooth and
featureless or one that has detectable features \cite{Cholis:2017ccs}.  

\subsubsection{Neutron star birth distribution}

As discussed in the Introduction, the cosmic-ray $e^{\pm}$
observed locally come from sources within a radius $R$ from us
where, as described in the Introduction, that radius is smaller
at higher energies. Thus, the observed cosmic-ray $e^\pm$ flux
is sensitive to the spatial distribution and birth rate within
this volume. The birth rate and spatial distribution of pulsars
within the Milky have been subjects of extensive work
\cite{Lorimer:2003qc,Lorimer:2006qs,FaucherGiguere:2005ny,Yusifov:2004fr}. Yet
there are great uncertainties in both, given the lack of a
complete pulsar survey of the sky at radio wavelengths.
Moreover, the pulsars' radio emission is highly anisotropic,
beamed with an opening angle spanning about one tenth of the
pulsars' $4\pi$ steradians. In fact, observations 
suggest that this ratio (typically referred to as the beaming
fraction) is time-dependent, being larger at the earlier stages
of the pulsar's evolution (as high as $\simeq 50\%$ 
during its first 10 kyr) and gradually decreasing
\cite{1998MNRAS.298..625T}. At gamma-ray wavelengths, 
the surveys do span the entirety of the sky but are sensitive
only to the brightest sources, i.e., the most powerful, younger,
and nearby members of the pulsar population.

The Milky Way pulsar birth rate has been estimated
to be $1.4\pm 0.2$ per century in Ref.~\cite{Lorimer:2006qs},
with alternative estimates that range between one and 
four per century at one $\sigma$ \cite{Vranesevic:2003tp,
FaucherGiguere:2005ny,Keane:2008jj} and even as high as $\simeq
8$ per century \cite{1999MNRAS.302..693D}. 
In our analysis, the pulsar birth rate is degenerate with the
fraction of spin-down power that goes to high energy $e^{\pm}$
and thus for simplicity we choose it to be one per century. 
The spatial distribution of pulsars at birth is expected to
follow the stellar distribution in the Milky Way's spiral arms.
It has been modeled in Refs.~\cite{FaucherGiguere:2005ny, 
Lorimer:2003qc, Lorimer:2006qs} based on the Parkes
multi-beam survey at 1.4 GHz \cite{Manchester:2001fp}. 
We generate simulations of Milky Way pulsar populations. 
To generate simulations of Milky Way pulsar populations, we 
follow both the parametrization of Ref.~\cite{Lorimer:2003qc}
and Ref.~\cite{Lorimer:2006qs} taking the latter as the
canonical distribution. More precisely, for the distribution 
of pulsars in galactocentric distance $r$ we use the radial density profile,
\begin{equation}
	\rho(r; B, C) = A \left(\frac{r}{R_{\odot}} \right)^B
	\exp \left( -C \left[ \frac{r - R_{\odot}}{R_{\odot}}\right] \right),
\end{equation}
where $B = 1.9$, $C=5.0$, $R_{\odot} = 8.5$ kpc, and $A$ is
normalized to a pulsar birth rate of one per
century. Furthermore, in our generated simulations, pulsars have
a distance $z$ away from the disk that follows a Laplace
distribution with a scale height of 50 pc and mean of 0 pc, in
accordance with  Ref.~\cite{FaucherGiguere:2005ny}. 
Finally, we do not try to simulate the spiral arms of the Galaxy, but 
simply assume a uniform distribution in Galactocentric angle. 

\subsubsection{Neutron-Star spin-down}

Neutron stars (NSs) are born from the core collapse of massive stars
in the range of 8--25 $M_{\odot}$. Given their violent birth combined with 
supernova explosions not being perfectly spherically
symmetric, neutron stars have large three-dimensional kick 
velocities (e.g. Ref.~\cite{Hobbs:2005yx} find kick velocity to
be $400 \pm 40$km/s) and also large ($\sim 10^{49}$ erg) initial
rotational energies. They also have strong magnetic fields due to
the contraction of the initial core, with large uncertainties
in the magnetic-field strengths due to
magnetohydrodynamic instabilities formed in the early
stages of the NS birth. The strength of the initial magnetic
fields at the poles ranges between $\sim10^{12}-10^{15}$
G. These rapidly rotating strong magnets will suffer the loss of
rotational energy with initial spin-down powers that may also span
orders of magnitude given the large uncertainties in the
initial magnetic  fields and rotational frequencies. This 
spin-down power evolves with time as,
\begin{equation}
     \dot{E}(t) = \dot{E_{0}}  \left(1 + \frac{t}{\tau_{0}}
     \right)^{-\frac{\kappa+1}{\kappa-1}}.
\label{eq:SpinDown}
\end{equation}
Here, $E_{0}$ is the initial rotational energy (i.e. $E_{0} =
1/2 \; I_{0} \Omega_{0}^{2}$, with $I_{0}$ the neutron-star
moment of inertia and $\Omega_{0}$ its initial angular
frequency),
\begin{equation}
     \tau_{0} = \frac{\dot{\Omega}}{(\kappa -1) \Omega} \left[1- \left(\frac{\Omega}
{\Omega_{0}}
\right)^{\kappa -1} \right]
\label{eq:Tau}
\end{equation}
is the characteristic timescale, or age, of a pulsar, and
$\kappa$ is the braking index describing the time evolution of
the neutron stars' angular frequency $\Omega$ through
$\dot{\Omega} \propto \Omega^{\kappa}$. Setting $\kappa = 3$
describes the spin-down due to magnetic-dipole radiation
\cite{1977puls.book.....M}. Measurement of $\kappa$
demands knowledge of $\Omega$, $\dot{\Omega}$ and
$\ddot{\Omega}$  ($\kappa \equiv \ddot{\Omega} \Omega /
\dot{\Omega}^{2}$). This biases the measurement toward
young pulsars where $\ddot{\Omega}$ is not too small to 
measure, and young pulsars may not be characteristic of the general
distribution.  Typical observed values give $\kappa \lsim 3$
\cite{1993MNRAS.265.1003L, 
1992ApJ...394..581G, Livingstone:2005jj, 1996Natur.381..497L, Camilo:2000mp, 
1994ApJ...422L..83K, Livingstone:2004fz, Espinoza:2016ars,
Marshall:2016kbp}, but 
there are also recent measurements of young pulsars with higher braking-index 
values \cite{Archibald:2016hxz}. Moreover the pulsar braking
index may evolve with time \cite{1977puls.book.....M,
1998MNRAS.298..625T} depending on the specific properties of the
pulsar \cite{Johnston:2017wgm}.

\begin{figure}
\hspace{0.0cm}
\includegraphics[width=3.5in,angle=0]{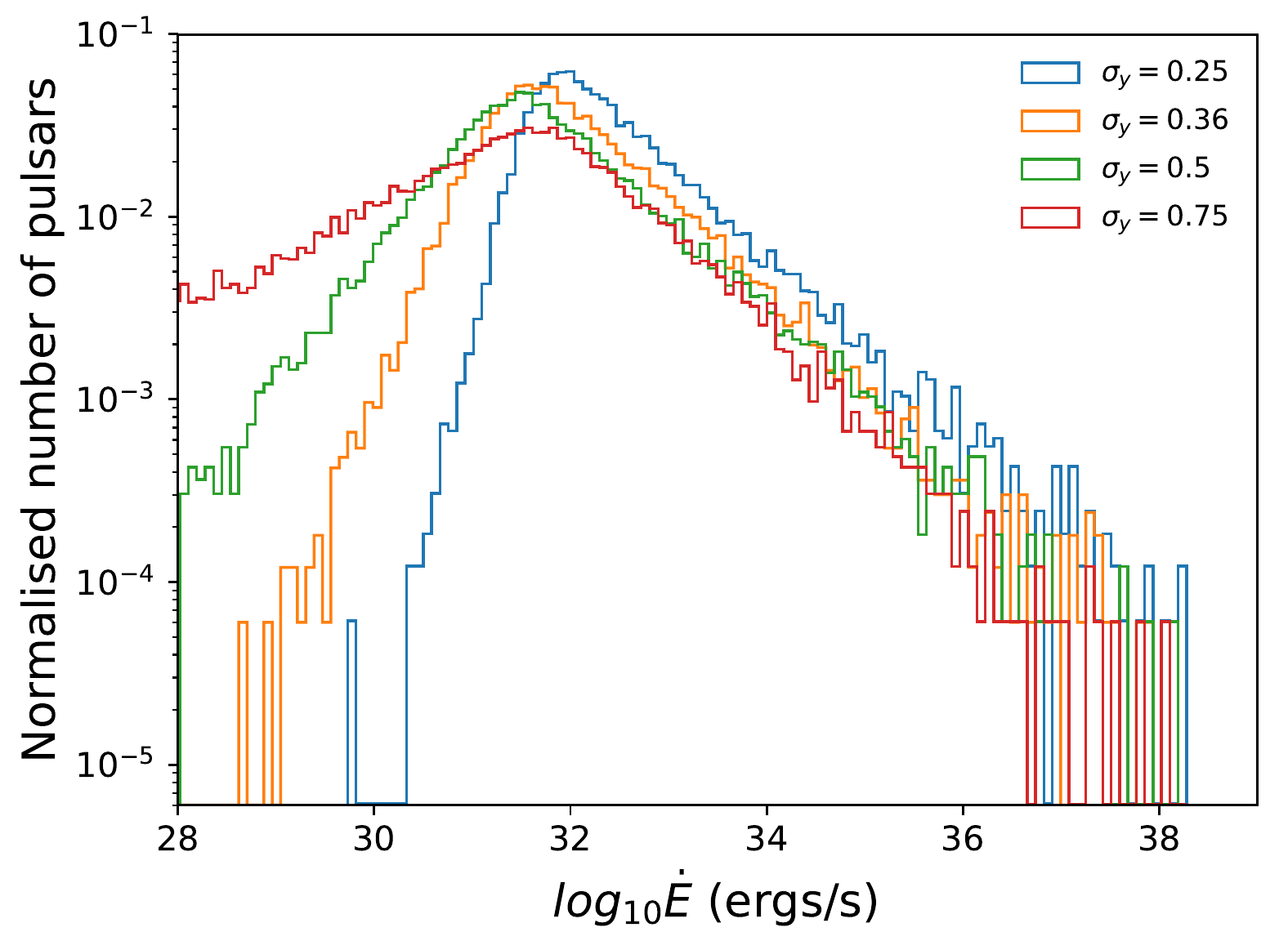}
\vspace{-0.4cm}
\caption{
     Shown here are normalized histograms of pulsar luminosities
     from various realizations. This is not their birth
     luminosity, but the luminosity evolved to today using
     Eq.~\ref{eq:SpinDown}. All simulations shown have a
     braking index of $\kappa = 3.0$.  The surface magnetic
     fields as well as periods of pulsars with distributions in
     orange and red are in good agreement with results shown in
     Fig. 6 of Ref.~\protect\cite{FaucherGiguere:2005ny}.  For
     the blue histogram, the surface magnetic fields of simulated
     pulsars and for the green, their periods agree well with
     Ref.~\protect\cite{FaucherGiguere:2005ny}. At high luminosities
     one can notice the Poisson fluctuations that arise in each
     individual simulation.  All distributions shown are
     associated with models that are allowed by the data.}
\label{fig:Luminosity_PDF}
\end{figure}

Given these uncertainties, our simulations test three
different choices, $\kappa=2.5$, 3.0, and 3.5 for the
braking-index. For each choice, we also choose a value for
the characteristic spin-down timescale $\tau_{0}$. Finally, we
account for pulsars not having a universal
initial spin-down power given the wide ranges of observed
magnetic fields for young pulsars ($\sim 10^{12}-10^{14.5}$
G). We simulate pulsars with an initial
spin-down power $\dot{E_{0}}$ given by
\begin{eqnarray}
     \dot{E}_0 &=& 10^{x} \textrm{ erg/s} =
     10^{x_{\textrm{cutoff}} - y} \textrm{ erg/s}, \nonumber 
     \textrm{ with} \\
     f(y) &=& \frac{\textrm{Exp}\left\{ - \frac{[- \mu_{y} +
     ln(y)]^{2}}{2 \sigma_{y}^{2}} \right\}} {\sqrt{2 \pi} y
     \sigma_{y}}.
\label{eq:InitialSpinDownPower}
\end{eqnarray}
The values for $x_{\textrm{cutoff}}$ and $\mu_{y}$ are
constrained by radio observations of rotation periods and modeled 
surface magnetic fields of Myr old pulsars \cite{FaucherGiguere:2005ny}.  
Moreover observation of the Crab
pulsar at  $\simeq 1$ kyr imposes a hard cut-off on observed spin-down 
power of pulsars larger than $10^{38.7}$ erg/s
\cite{Manchester:2004bp, ATNFSite}. In our simulations we take
$\sigma_{y} = [0.25, 0.36, 0.5, 0.75]$. Those observational
constraints result in pulsars with $\tau_{0}$ values as small as
0.6 kyr for $\kappa = 2.5$ and as large as 30 kyr for $\kappa =
3.5$, with typical values of 6--10 kyr for breaking index of
3.0.  In Fig.~\ref{fig:Luminosity_PDF}, we show normalized
histograms of $\dot{E}$ for each value of $\sigma_y$ for
simulations that are allowed by the data.

We allow for a wide range of assumptions regarding the true
underlying current period of pulsars with ages $\leq 10$ Myr,
as well as their surface magnetic fields. Since we rely on
observations of pulsars with ages of order $10^{5}$-$10^{7}$
years, we probe predominantly that population and not the
spin-down conditions in the very early stages. We then 
use the CR measurements to constrain the birth properties of the
pulsar population.

We also note that neutron-star kick velocities of $\sim100$ km/s
result in a displacement of $\sim100$ pc of the
NSs from their birth location within 1 Myr, but only a few pc in
their first $\sim$10 kyr ($\simeq \tau_{0}$) that is relevant
for our work.  We thus take their distribution in
space to be their distribution at birth.

\subsubsection{Injection properties of cosmic-ray $e^{\pm}$}

From radio and microwave observations of synchrotron radiation 
close to the NS magnetic poles and of
inverse Compton scattering in gamma rays further away from the
NS, we know that $e^\pm$ pairs are produced and get accelerated
inside the pulsar magnetospheres (up to distance scales
$O(10^{4} -  10^{5})$ km) \cite{Rees:1974nr, Arons:1979bc,
Cheng:1986qt, Daugherty:1995zy, 
Contopoulos:1999ga, Komissarov:2005xc, Gruzinov:2004jc, Contopoulos:2005rs, 
Spitkovsky:2006np, Harding:2008kk, Kalapotharakos:2008zc, Watters:2008bp, 
Bai:2009wg, Buehler:2013yqa, Cerutti:2014ysa}. In addition,
$e^{\pm}$ can get accelerated outside the pulsar magnetosphere
before or at the PWN termination shock that typically extends
out to $\sim$pc distances from the NS \cite{Goldreich:1969sb,
Hoshino:1992zz, Lyubarsky:2000yn, Lyubarsky:2003wv,
Sironi:2011zf, Buehler:2013yqa, 
Sironi:2014jfa,  Zenitani:2014hea, Kargaltsev:2015pma}. In fact,
there is evidence for the presence of $\sim 100$ TeV $e^\pm$
at even larger distances, of $\simeq 10$ pc,
from HAWC observations of gamma rays at $\sim10$ TeV from the 
pulsars Geminga and Monogem \cite{Abeysekara:2017hyn,
Abeysekara:2017old}.  All these observations suggest that
pulsars are environments that are rich in high-energy $e^{\pm}$,
a fraction of which may escape into the ISM as
cosmic rays.  

Following Ref.~\cite{Malyshev:2009tw}, we assume that each
pulsar is a point source of CR $e^{\pm}$ described by a source
term,
\begin{eqnarray}
     Q(E,\vec{x},t) &=& \frac{dN}{dE}\delta(\vec{x}) 
     \left(1 + \frac{t}{\tau_{0}}
     \right)^{-\frac{\kappa+1}{\kappa-1}}, \; \textrm{with}
     \nonumber \\ 
     \frac{dN}{dE} &=& Q_{0} \big(\frac{E}{1
     \textrm{GeV}}\big)^{-n} Exp\{-\frac{E}{E_{\textrm{cut}}}
     \},
\label{eq:PulsarSource}
\end{eqnarray}
the CR energy density from a given pulsar. Here, $\delta(\vec
x)$ is a Dirac delta function localized at the pulsar position,
and the normalization of $Q_{0}$ is such
that~\cite{Malyshev:2009tw},
\begin{equation}
     \int \int \int dE \, d\vec{x} \, dt \, Q(E,\vec{x},t) =
     \eta W_{0},
\label{eq:Qnorm}
\end{equation}
where $\eta$ is the fraction of the rotational energy $W_{0}$ that
has already been lost through CR $e^{\pm}$ injected into the ISM. This
gives,
\begin{equation}
     Q_{0} = \frac{E_{\textrm{cut}}^{2-n}}{\Gamma(2-n)} \,
     \frac{2}{(\kappa -1) \tau_{0}},
\label{eq:Q0}
\end{equation}
where $\Gamma(2-n)$ is the Euler gamma-function, $n<2$, and
$\kappa>1$. The total amount of available rotational energy
depends on the exact initial spin-down power and its time
evolution. We use Eq.~\ref{eq:SpinDown}, which for $t \gg \tau_{0}$,
gives $W_{0}  = \dot{E_{0}} \tau_{0}$, while for $t \sim \tau_{0}$,
there is a correction factor of $(1 + t/\tau_{0})^{(\kappa
+1)/(\kappa - 1)}$.

We are agnostic on the exact values of $\eta$ and $n$ of each
pulsar, but X-ray and $\gamma$-ray observations
suggest values of $n\simeq 1.4-2.0$
\cite{Bietenholz:1997xz, 1993ApJ...415..286H,
1998ApJ...494..734F, Thompson:1998ct, 
Atoyan:1999vi, 2001A&A...378..918K, Abdo:2009ec, etc.:2010vw, 
TheFermi-LAT:2013ssa}, even though observations of gamma rays
from the Crab pulsar reveal a significantly softer spectrum for
the high-energy CR $e^{\pm}$\cite{1998ApJ...494..734F,
Abdo:2009ec}. There are thus significant observed
source-to-source variations among pulsars, especially at
higher energies. In our simulations, we do not assume that all
pulsars have the same values of CR $e^{\pm}$ injection 
indexes and spin-down power efficiencies (to CRs). Instead, the
parameter $n$ follows a uniform distribution $g(n)$.  We use two
different assumptions for $g(n)$ in assigning an injection index
$n$ to each pulsar of a given simulation; they are
\begin{eqnarray}
     n \epsilon \begin{cases} [1.6,1.7], \,\,\,\,\,\,\,
     \textrm{option "a"} \\  
     [1.4,1.9], \,\,\,\,\,\,\, \textrm{option "b"}.
     \end{cases}
\label{eq:gn}
\end{eqnarray}

Similarly, each pulsar in a given simulation has a value
of $\eta$ taken from a log-normal distribution
\cite{Cholis:2017ccs},
\begin{equation}
     h(\eta) = \frac{Exp \left\{ -\frac{ \left[ - \mu +ln(-1 +
     \eta) \right]^{2}}{2 \sigma^2}\right\}} 
     {\sqrt{2 \pi} (\eta -1) \sigma}.
\label{eq:eta}
\end{equation}
We do not know what the exact range of the $\eta$ values should
be. We therefore have in our simulations three options in
choosing values for $\mu$ and $\sigma$ in 
Eq.~\ref{eq:eta}. 
The more physically intuitive quantities are
the mean efficiency $\bar{\eta} = 1 + \textrm{Exp} \left\{ \mu +
\frac{\sigma^2}{2} \right\}$ and the parameter $\zeta =
10^{\sqrt{\sigma}}$.  For a given produced simulation, $\zeta$ is
fixed but $\bar{\eta}$ is normalized to the CR data. The three
choices for log-normal distributions $h(\eta)$ are 
\begin{eqnarray}
      (\bar{\eta}, \zeta) = \begin{cases} (2\times 10^{-2},
      1.29), \,\,\,\,\,\,\, \textrm{option "a"} \\  
      (4\times 10^{-3}, 1.47), \,\,\,\,\,\,\, \textrm{option
      "b"} \\
      (1\times 10^{-3}, 2.85), \,\,\,\,\,\,\, \textrm{option "c"},
      \end{cases}
\label{eq:hn}
\end{eqnarray}
where the $\bar{\eta}$ values refer to the starting point before
the fit. Typically, the values of $\bar{\eta}$ do not change by
more than a factor of a few, with large values of $\zeta$
leading to smaller values of $\bar{\eta}$. 

We note that the exact assumption for the value of the injection
upper cutoff $E_{\textrm{cut}}$ does not affect our results as
$e^{\pm}$ that propagate into the ISM cool down very rapidly.
We take $E_{\textrm{cut}} = 10$ TeV.

\subsubsection{Propagation of Cosmic rays} 

Cosmic rays injected into the ISM by individual pulsars have to
travel from their simulated locations in the Milky Way to
the Earth's location where they are detected by the
\textit{AMS-02}, \textit{DAMPE}, and \textit{CALET}
instruments. They propagate first through the ISM before
entering the volume affected by the solar magnetic field and
wind. During that first propagation, the $e^{\pm}$ diffuse
through the complicated galactic magnetic-field structure and lose
energy via synchrotron radiation as well as ICS with the CMB and
ambient infrared and optical photons. The first process
is described by the diffusion coefficient,
\begin{equation}
     D(E) = D_{0} (E/1 \textrm{GeV)}^{\delta},
\label{eq:diff}
\end{equation} 
assumed for simplicity to be homogeneous and isotropic within a
thick several-kpc diffusion disc around the Milky Way stellar
disc, where the pulsars reside. The energy losses are described by
\begin{equation}
     dE/dt = - b \, E^{2},
\label{eq:Eloss}
\end{equation} 
where $b$ is proportional to the sum of the energy densities of
the local (within a few kpc) magnetic field $U_{\textrm{B}}$ and
the local radiation field $U_{\textrm{rad}}$. Relying on
previous studies of the CR boron-to-carbon (B/C) ratio observed
by both \textit{PAMELA} and \textit{AMS-02} and also on the CR
proton data observed by \textit{PAMELA} and \textit{Voyager 1},
\cite{Malyshev:2009tw, Trotta:2010mx, Cholis:2015gna} we adopt
five different ISM models for the CR propagation in our analysis. 
These are described in Table~\ref{tab:ISMBack}.

\begin{table}[t]
    \begin{tabular}{cccc}
         \hline
           Model & $b$ ($\times 10^{-6}$GeV$^{-1}$kyr$^{-1}$) & $D_{0}$ (pc$^2$/kyr) & $\delta$\\
            \hline \hline
            A1 &  5.05 & 123.4 & 0.33 \\
            C1 &  5.05 & 92.1 & 0.40 \\
            C2 &  8.02 & 92.1 & 0.40 \\
            C3 &  2.97 & 92.1 & 0.40 \\      
            E1 &  5.05 & 58.9 & 0.50 \\
        \hline \hline 
        \end{tabular}
\caption{The parameters for each of our five ISM
propagation models; see Eqs~\ref{eq:diff},~\ref{eq:Eloss} and
text for details.} 
\label{tab:ISMBack}
\end{table}

After their ISM propagation and before their detection, CRs
travel through the time-evolving solar wind. For energies
$E \leq 10$ GeV, the effect on the CR spectra
is known as solar modulation. This is the imprint of the diffusion, 
drift, and adiabatic energy losses experienced by CRs
traveling through the complex magnetic field structure of the
heliosphere (within $\simeq$100 au from the Sun).  This solar
modulation is described by imposing a translation in the energy
of the CR spectra as 
\cite{1968ApJ...154.1011G},
\begin{eqnarray}
     \frac{dN^{\oplus}}{dE_{\textrm{kin}}} (E_{\textrm{kin}})
     &=& \frac{(E_{\textrm{kin}}+m)^{2} -m^{2}} 
     {(E_{\textrm{kin}}+m+| Z| e \Phi)^{2} -m^2} \nonumber \\ 
     &\times&\; \frac{dN^{\rm ISM}}{dE_{\textrm{kin}}}
     (E_{\textrm{kin}}+| Z| e \Phi).
\label{eq:SolMod}
\end{eqnarray}
Here, $E_{\textrm{kin}}$ is the observed kinetic CR energy at
Earth ($\oplus$), while $\frac{dN^{\oplus
(\textrm{ISM})}}{dE_{\textrm{kin}}}$ the differential CR flux at
Earth. The equivalent $E_{\textrm{kin}}$ for the ISM
spectrum is (on average) $E_{\textrm{kin}} + | Z| e \Phi$,
where $\Phi$ is the modulation potential.  Finally,
$| Z| e$  is the absolute value of the CR charge.

Ref.~\cite{Cholis:2015gna} used archival data to obtain a
time-, charge- and  rigidity(R)-dependent formula
\begin{eqnarray}
     \Phi(R,q,t) = &\phi_{0}& \, \bigg( \frac{|B_{\rm
     tot}(t)|}{4\, {\rm nT}}\bigg) + \phi_{1} \, H(-qA(t))\,  
     \bigg( \frac{|B_{\rm tot}(t)|}{4\,  {\rm nT}}\bigg)
     \nonumber \\ 
     &\times& \,\bigg(\frac{1+(R/R_0)^2}{\beta
     (R/R_{0})^3}\bigg) \, \bigg( \frac{\alpha(t)}{\pi/2}
     \bigg)^{4}
\label{eq:ModPot}
\end{eqnarray}
for the solar modulation potential.
Following that work, $R_{0}$ is set to 0.5 GV. Instead
$\phi_{0}$ and $ \phi_{1}$ are marginalized within $[0.32,0.38]$
GV and $[0,16]$ GV respectively. The values for
$B_{\textrm{tot}}(t)$ and $\alpha(t)$ are averaged over 6-month
periods using the measurement by the \textit{ACE} satellite
\cite{ACESite} and the models of the Wilcox Solar Observatory
\cite{WSOSite}. The potential $\Phi(R,q,t)$ is time dependent for every
species, different between electrons and positrons observed at
the same time and rigidity, and is smaller for larger
rigidities becoming $\Phi(R,q,t) \rightarrow \phi_{0}
(|B_{\textrm{tot}}(t)|)/(4\, \textrm{nT})$. 

\subsubsection{Combining all the pulsar-population uncertainties} 

To combine all the astrophysical uncertainties, we first
generate a pulsar population in the Milky Way with an assumed
spatial distribution and birth rate, and a given choice of
$\kappa$, $\tau_{0}$, and $f(y)$ characterizing the initial
spin-down power and its evolution. There are 30 different
combinations of these assumptions 
(see Appendix A of \cite{Cholis:2017ccs} for details). These
pulsars also follow the distributions $g(n)$ and $h(\eta)$ 
on the injection properties.  There are six 
combinations for these two distributions, which we refer
to as "aa", "ba", "ab", "bb", "ac",  and "bc",
where the first letter refers to the $g(n)$ options "a" or "b"
of Eq.~\ref{eq:gn} and the second letter to $h(\eta)$, with its
options "a-c" of Eq.~\ref{eq:hn}.  We also choose one of the
five ISM propagation models described by
Table~\ref{tab:ISMBack}, while we marginalize over parameters 
$\phi_{0}$ and $\phi_{1}$ of Eq.~\ref{eq:ModPot} to account for
solar modulation. Those choices result in $30 \times 6 \times 5
= 900$ different astrophysical realizations/simulations that we
test.

\begin{figure}
\hspace{-0.0cm}
\includegraphics[width=3.7in,angle=0]{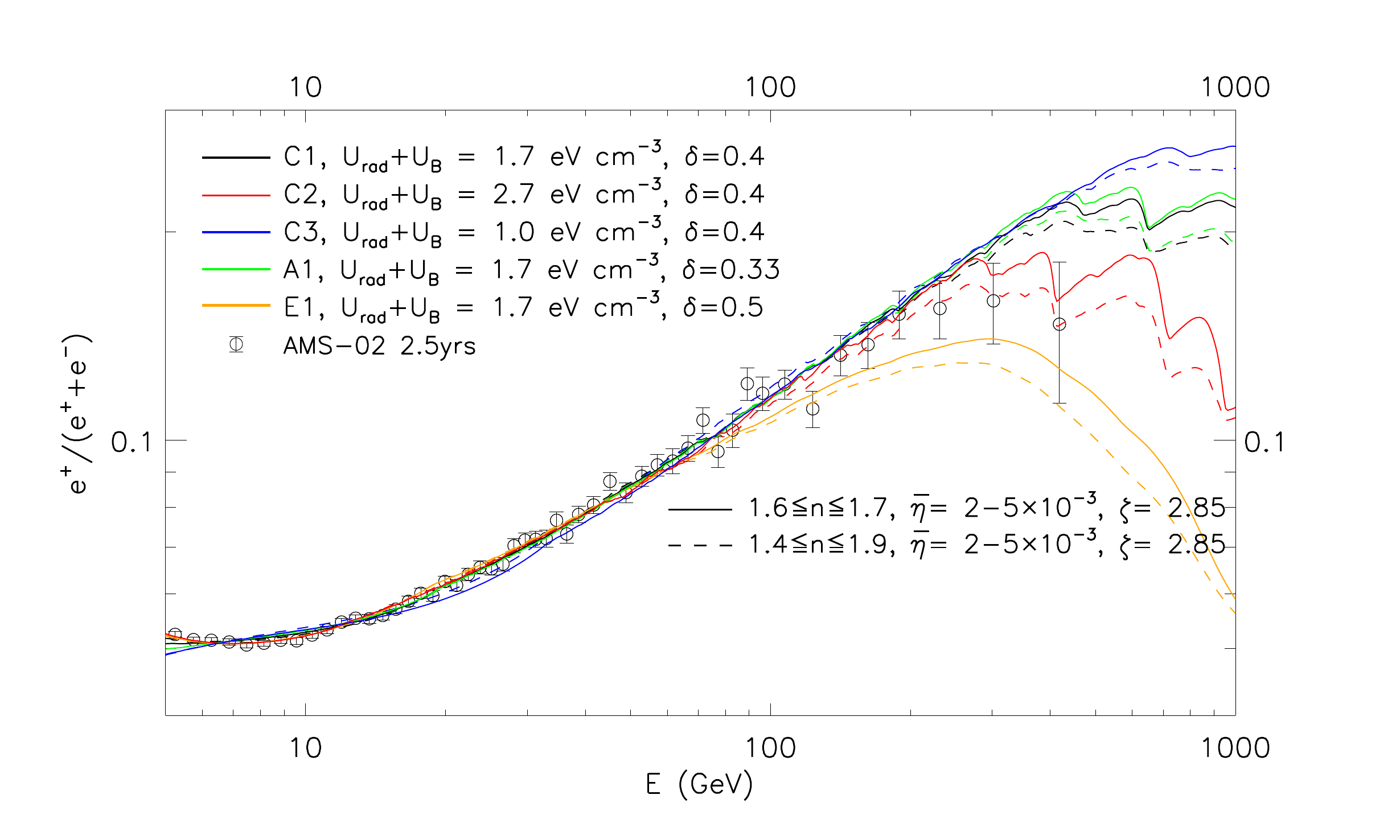}\\
\vspace{-0.3cm}
\includegraphics[width=3.7in,angle=0]{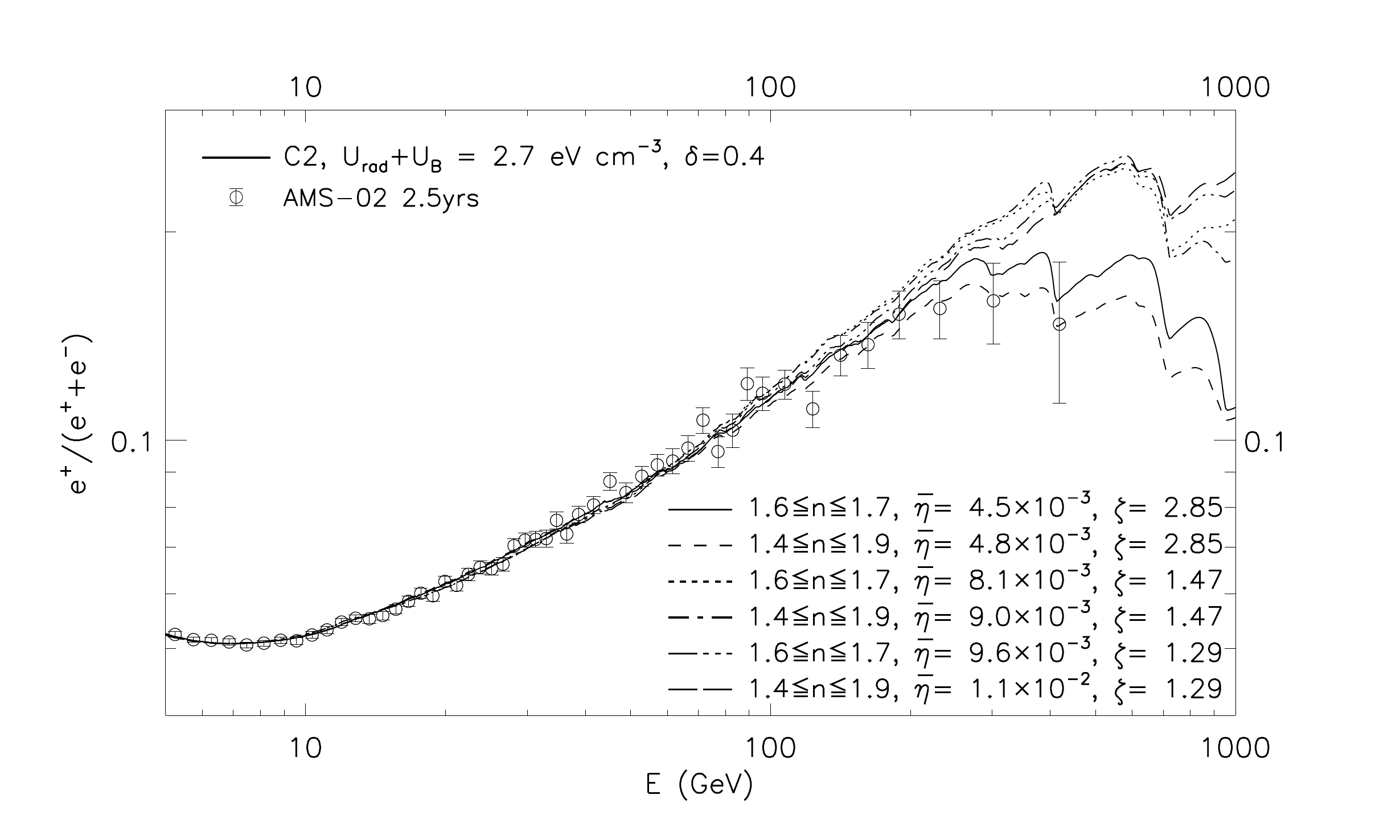}
\vspace{-0.6cm}
\caption{The effects of changing assumptions in our
     astrophysical pulsars realizations. \textit{Top:} Different
     color lines depict the effect that changing only the CR
     propagation through the ISM, (i.e. CR diffusion and
     energy-losses described by Eqs.~\ref{eq:diff}
     and~\ref{eq:Eloss}) has on the expected flux from what is
     otherwise the same simulation of pulsar sources in the
     Milky Way. ISM Models "C1", "C2", "C3", "A1" and "E1" are
     described in Table~\ref{tab:ISMBack}. As described in the
     main text, we fit the simulations to the \textit{AMS-02}
     data. The solid vs dashed lines for every given colored
     line show  the impact on our results of changing solely the
     distribution of the CR injection indexes $g(n)$ (see
     Eq.~\ref{eq:gn} and also
     Eq.~\ref{eq:PulsarSource}). \textit{Bottom:} For a specific
     assumption on the ISM propagation conditions ("C2"), we
     show in the six types of lines, the effect of varying the
     assumptions on $g(n)$ and $h(\eta)$ (see
     Eqs~\ref{eq:PulsarSource},~\ref{eq:gn}
     and~\ref{eq:eta},~\ref{eq:hn} respectively and discussion
     therein).}
\label{fig:AstroRealiz}
\end{figure}

Each one of these astrophysical realizations is fitted to the CR
data by allowing for five fitting parameters, $\phi_{0}$,
$\phi_{1}$, the normalization of the primary CR electron  flux,
the normalization of the secondary CR $e^{\pm}$ fluxes and the
normalization of the  total pulsar CR $e^{\pm}$ fluxes at the
location of the Sun (outside the heliosphere).  The impact of
these different astrophysical choices on the positron fraction
spectrum is given in Fig.~\ref{fig:AstroRealiz}. The five different 
colored lines refer to the five ISM models. 

In the top panel of Fig.~\ref{fig:AstroRealiz} and for the
choice "c" of $h(\eta)$, we give the two choices for $g(n)$;
i.e., the choices "ac" and "bc" in solid and dashed lines  
respectively. Different ISM assumptions can enhance ("A1", "C2"
choices) or suppress ("C3", "E1" choices) the small-scale (in
energy range) features of the spectra. At high energies when the
energy-loss rate is assumed to be larger (smaller) or the CR
diffusion slower (faster), the small-scale spectral features are
more (less) pronounced. Similarly, when a wider range of
injection indexes $n$ is assumed, the resulting features are
more evident (from the fact that Poisson fluctuations of
nearby sources with hard injected spectra being dominant are
more common). Moreover, since the fits are dominated by the
low-energy data, different ISM models can predict significant
variations at high energies. For instance, "A1" models faster
diffusion of low-energy CRs, but with  
a smaller diffusion index $\delta =0.33$, which in turn increases
the escape timescale of CRs from the Galaxy, and thus 
enhancing their flux at high energies. 
Model "E1" represents the reverse
assumption (slow diffusion of low-energy CRs but with a larger 
diffusion index $\delta = 0.5$ leading to faster escape at high
energies) suppressing the high-energy fluxes. Model "C1"
represents a more intermediate case. Also, larger/smaller energy
losses (modeled by "C2"/"C3") suppress/enhance the high-energy
CR pulsar fluxes.

The bottom panel of Fig.~\ref{fig:AstroRealiz} shows, for a fixed
choice  ("C2") of ISM assumptions, the impact of varying the
$g(n)$ and $h(\eta)$ choices. A wider range of the $\zeta$
parameter, associated with the standard deviation in the
$h(\eta)$, results in more pronounced features. We also note
that when $\zeta$ is larger, a few pulsars may deposit a larger
fraction of their spin-down power to CR $e^{\pm}$ in the ISM. In
turn, the fits are forced to compensate for that by reducing the
averaged $\bar{\eta}$ value.

\section{Results}
\label{sec:results}

\subsection{Using only the \textit{AMS-02} positron fraction data}
\label{sec:AMSfit}

\begin{figure}
\vspace{0.45cm}
\hspace{-0.4cm}
\includegraphics[width=3.45in,angle=0]{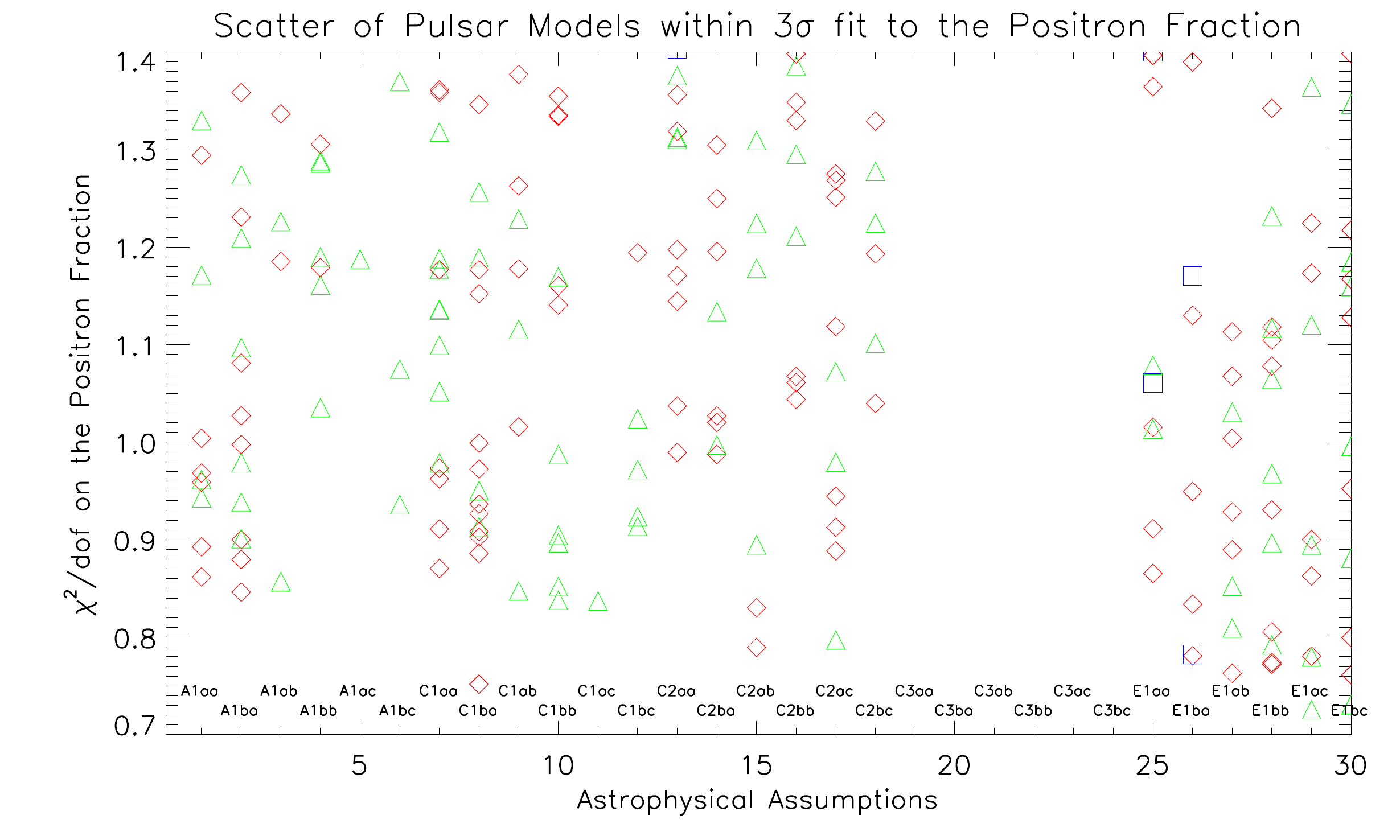}
\vspace{-0.0cm}
\caption{The scatter of astrophysical pulsar realizations that fall within our 3$\sigma$ 
constraint to the \textit{AMS-02} positron fraction data. On the y-axis we give the 
$\chi^{2}$/d.o.f. value, while on the x-axis we give the astrophysical assumptions, 
e.g. "A1aa" refers to ISM model "A1" and $g(n)$ and $h(\eta)$ distributions choice 
"aa" of Eqs.~\ref{eq:gn} and~\ref{eq:hn} respectively. Blue boxes are for braking index 
$\kappa$ of 2.5, red diamonds for $\kappa=3.0$ and green triangles for $\kappa=3.5$. 
Two evident results are that i) very few pulsar population realizations with $\kappa=2.5$ 
survive the positron fraction fit and ii) there are no realizations with ISM conditions "C3" 
allowed by the data (see text for further details).} 
\label{fig:AMS2014scatter}
\end{figure}

Our fits to the \textit{AMS-02} positron-fraction spectrum allow us to constrain 
combinations of the above mentioned astrophysical uncertainties via data not previously 
used to probe the pulsar population properties of the Milky Way. Of the 900 astrophysical 
realizations, only 205(160) can fit the positron fraction spectrum within 3$\sigma (2\sigma$)
 from an expectation of $\chi^{2}$ of 1 for each degree of freedom \footnote{There are 
 51 energy bins for the positron fraction in the energy range of 5 to 500 GeV that we 
 fit. Since in the fitting, we have five free parameters (as discussed 
Section~\ref{sec:method}), there are 46 degrees of freedom leading to a total $\chi^{2}$ 
of 64.2 and 57.3 respectively for the quoted  3$\sigma$ and 2$\sigma$ ranges assuming 
as a starting point a $\chi^{2}$/d.o.f.=1.} These astrophysical realizations are depicted 
in Fig.~\ref{fig:AMS2014scatter}. 

In Fig.~\ref{fig:AMS2014scatter}, the y-axis gives the $\chi^{2}$ per degree of freedom 
for the positron fraction data. The x-axis represents the probed parameter space in 
30 distinct combinations of assumptions. These mark the assumptions on the CR 
$e^{\pm}$ ISM propagation described by the first two characters ("A1", "C1", "C2", "C3", 
"E1") as well as the CR injection properties both with regards to the injection index 
$n$-distribution $g(n)$ of Eq.~\ref{eq:gn} and the $h(\eta)$ distribution of Eq.~\ref{eq:hn}. 
The latter assumptions are depicted by the last two characters ("aa", "ba", "ab", "bb", 
"ac", "bc"). For example the first discrete point in the x-axis ("A1aa") refers to ISM model 
"A1" of Table~\ref{tab:ISMBack} while the third character "a" stands for $n \epsilon 
[1.6,1.7]$ and the last character for $(\bar{\eta}, \zeta) = (2\times 10^{-2}, 1.29)$ 
(case "a"). Each allowed pulsar population simulation is represented by a shape. Blue 
boxes are for $\kappa = 2.5$, red diamonds are for $\kappa = 3.0$ and green triangles 
for $\kappa = 3.5$.

From our fits to the positron fraction, we have two major findings.  The first is that 
while we run 240 realizations with a braking index of $\kappa=2.5$, only 5(3) survive within 
the 3(2) $\sigma$ fit threshold. While there are constraints from radio 
observations on the period and NS surface B-field, we try to adopt as wide as possible 
assumptions on the initial spin-down properties of the pulsar populations. These wide
assumptions have a significant impact also on the observed pulsar populations as 
we show in Fig.~\ref{fig:Luminosity_PDF}. For the braking index value of 2.5 
assumed in the 240 simulations, values for $\tau_{0}$ are in the range of 0.6-1 kyr and of
$10^{x_{\textrm{cutoff}}}$ from $10^{38.2}-10^{39}$ erg/s with varying assumptions on 
$\mu_{y}$ and $\sigma_{y}$ (see Eqs.~\ref{eq:SpinDown} 
and~\ref{eq:InitialSpinDownPower} and Appendix B of \cite{Cholis:2017ccs}).   
Our second major finding relates to the ISM conditions. For every choice of NS 
population birth distribution, spin-down properties and pulsars injection of CR 
$e^{\pm}$ properties, we test each of our five ISM models. Our model "C3" is always 
excluded as can be seen by the gap in the parameter space in Fig.~\ref{fig:AMS2014scatter}.
Table~\ref{tab:Kappa_VS_ISM} describes the above and provides the fraction 
of simulations that are allowed by the data (at 3$\sigma$) with the exact number of
simulations given in the parentheses.
\begin{table}[t]
    \begin{tabular}{c|ccccc}
         \hline
             & A1 & C1 & C2 & C3 & E1 \\
            \hline \hline
            $\kappa = 2.5$ & 0 (0) & 0 (0) & 0.02 (1) & 0 (0) & 0.08 (4) \\
            $\kappa = 3.0$ & 0.27 (21) & 0.40 (31) & 0.40 (31) & 0 (0) & 0.46 (36) \\
            $\kappa = 3.5$ & 0.37 (20) & 0.43 (23) & 0.35 (19) & 0 (0) & 0.41 (23) \\
            \hline \hline 
        \end{tabular}
\caption{For combinations of the three choices of braking index $\kappa = 2.5, 3.0, 3.5$ 
and the five choices of ISM propagation conditions "A1", "C1", "C2", "C3", "E1", we give 
the fraction of pulsar population simulations that are consistent (within our 3$\sigma$ 
threshold) to the \textit{AMS-02} positron fraction spectrum. In the parentheses we give 
the actual number of the simulations allowed. For the combination of $\kappa=3$ and 
"A1" we produced 78 simulations to probe the remaining astrophysical parameters, of 
which 21 (27$\%$) are allowed.}
\vspace{-0.2cm}
\label{tab:Kappa_VS_ISM}
\end{table}

Both sets of assumptions (i.e. $\kappa = 2.5$ or ISM model "C3") systematically predict 
a higher positron fraction than observed above 100 GeV. For the case of  the ISM 
"C3" model, this is straightforward given the suppressed energy-loss coefficient $b$ 
of Eq.~\ref{eq:Eloss}. For $\kappa=2.5$, the explanation is as follows. Smaller braking 
index values demand not only a smaller characteristic timescale $\tau_{0}$, but also 
result in a faster spinning-down of the pulsars at $t \gg \tau_{0}$, e.g $\dot{E}(t) 
\propto t^{-2.33}$ for $\kappa=2.5$ vs $\dot{E}(t) \propto t^{-2}$ for $\kappa=3$, 
(see Eq.~\ref{eq:SpinDown}). In setting the spin-down choices of our pulsar population 
simulations, we rely on radio observations. These observations probe a wide range of 
pulsar ages, but there are many more pulsars of age 1-10 Myr than of 10-100 kyr. In 
our simulations, we check that our pulsar populations are consistent with the late-age 
properties of the observed pulsars and then effectively evolve backwards in time the 
pulsars spin-down. Yet, as we explained in section~\ref{sec:method}, most of the pulsars 
CR flux is produced in the early stages of their evolution.  A more abrupt spin-down-power 
time-evolution as in the $\dot{E}(t) \propto t^{-2.33}$ vs $\propto t^{-2}$ case predicts 
a higher CR flux overall from each pulsar at a level that is already inconsistent with the 
CR \textit{AMS-02} data. Values of $\kappa < 2.5$, as are observed for 
some of the youngest pulsars, are excluded. Hence, CR data also suggest a time evolution of 
the braking index. 

Moreover, we note that changing the braking index to $\kappa \geq 3$ is enough 
to exclude the tension with the positron fraction data. Yet as more events are detected  
from \textit{AMS-02} and if indeed a 
cut-off / drop-off the positron fraction above 500 GeV is confirmed, we will be able to 
probe/exclude more simulations with $\kappa = 3$. The ISM model "A1" that predicts 
higher CR fluxes due to its diffusion parameters (see Table~\ref{tab:ISMBack} and 
Figure~\ref{fig:AstroRealiz} \textit{top}) is currently only mildly less preferred as shown 
also in Table~\ref{tab:Kappa_VS_ISM}, but we do find that with an observed drop-off of
the positron fraction above 500 GeV, many pulsar simulations with the "A1" 
assumptions can be excluded, while ISM model assumptions "E1" will be slightly 
preferred ("C1" and "C2" being less affected by the presence of the claimed cut-off). 

In addition to the above findings, we note a preference among astrophysical realizations 
that predict a narrower distribution $h(\eta)$, in the fraction of the spin-down power that goes into 
CR $e^{\pm}$, with $\zeta$ following hierarchically option "a" 
($\zeta \equiv 10^{\sqrt{\sigma}} = 1.29$) over option "b" ($\zeta = 1.47$) over option "c" 
($\zeta = 2.85$) as described by Eqs.~\ref{eq:eta} and~\ref{eq:hn}. This can be seen 
more directly by our results in Table~\ref{tab:Injec_VS_ISM}, where we show the slice 
in parameter space describing the combination of $g(n)$ and $h(\eta)$ injection 
properties vs the five ISM propagation models. For the narrower $h(\eta)$, anywhere 
between 23$\%$ and 50$\%$ of our simulations are within our threshold of fit 
to the positron fraction data. For the wider distribution choices, these fractions are reduced 
with the widest choice (choices "ac" and "bc") having anywhere between 7$\%$ to 
43$\%$ of the simulations within our 3$\sigma$ fitting threshold.
\begin{table*}[t]
    \begin{tabular}{c|c|ccccc}
         \hline
             & & A1 & C1 & C2 & C3 & E1 \\
            \hline \hline
            aa & $1.6 \leq n \leq 1.7$, $\eta = 2\times 10^{-2}$, $\zeta = 1.29$   & 0.37 (11) & 0.43 (13) & 0.36 (11) & 0 (0) & 0.33 (10) \\
            ba & $1.4 \leq n \leq 1.9$, $\eta = 2\times 10^{-2}$, $\zeta = 1.29$   & 0.46 (14) & 0.50 (15) & 0.27 (8) & 0 (0) & 0.23 (7) \\
            ab & $1.6 \leq n \leq 1.7$, $\eta = 4\times 10^{-3}$, $\zeta = 1.47$   & 0.13 (4) & 0.23 (7) & 0.27 (8) & 0 (0) & 0.33 (10) \\
            bb & $1.4 \leq n \leq 1.9$, $\eta = 4\times 10^{-3}$, $\zeta = 1.47$   & 0.23 (7) & 0.36 (11) & 0.33 (10) & 0 (0) & 0.43 (13) \\
            ac & $1.6 \leq n \leq 1.7$, $\eta = 1\times 10^{-3}$, $\zeta = 2.85$   & 0.07 (2) & 0.07 (2) & 0.27 (8) & 0 (0) & 0.33 (10) \\
            bc & $1.4 \leq n \leq 1.9$, $\eta = 1\times 10^{-3}$, $\zeta = 2.85$   & 0.10 (3) & 0.17 (5) & 0.20 (6) & 0 (0) & 0.43 (13) \\
            \hline \hline 
        \end{tabular}
\caption{As in Table~\ref{tab:Kappa_VS_ISM}, we present, for the
     combination of the six choices ("aa", "ba", "ab", "bb", "ac" and "bc")
     of $g(n)$ and $h(\eta)$ and the five ISM models, the
     fraction of pulsar simulations that are consistent within
     3$\sigma$ with the \textit{AMS-02} positron fraction. The
     numbers in parentheses give the number of the simulations
     allowed.} 
\vspace{-0.1cm}
\label{tab:Injec_VS_ISM}
\end{table*}

We have found no preference towards a narrower or a wider
distribution for $\dot{E}_{0}$---i.e., for the parameter
$\sigma_{y}$ of Eq.~\ref{eq:InitialSpinDownPower}.  

While still more data from CRs will be necessary, the preference
towards a narrow $h(\eta)$ suggests that the pulsar environments
do not have very wide source-to-source variations with respect
to their output of CR $e^{\pm}$ injected into the ISM.

Finally, we find that there is only a slight preference for the
wider $g(n)$ injection index $n$-distribution (option "b" of
Eq.~\ref{eq:gn}) over the narrower one (option "a" of
Eq.~\ref{eq:gn}). This is evident in  
the slice of parameter space in Table~\ref{tab:Injec_VS_Kappa}
showing the $g(n)$, $h(\eta)$ parameters versus the braking 
index $\kappa$.

\begin{table*}[t]
    \begin{tabular}{c|c|ccc}
         \hline
             & & $\kappa = 2.5$ & $\kappa = 3.0$ & $\kappa = 3.5$ \\
            \hline \hline
            aa & $1.6 \leq n \leq 1.7$, $\eta = 2\times 10^{-2}$, $\zeta = 1.29$ & 0.08 (3) & 0.38 (25) & 0.36 (16)  \\
            ba & $1.4 \leq n \leq 1.9$, $\eta = 2\times 10^{-2}$, $\zeta = 1.29$ & 0.05 (2) & 0.48 (31) & 0.27 (12)  \\
            ab & $1.6 \leq n \leq 1.7$, $\eta = 4\times 10^{-3}$, $\zeta = 1.47$ & 0 (0) & 0.26 (17) & 0.27 (12)  \\
            bb & $1.4 \leq n \leq 1.9$, $\eta = 4\times 10^{-3}$, $\zeta = 1.47$ & 0 (0) & 0.32 (21) & 0.44 (20) \\
            ac & $1.6 \leq n \leq 1.7$, $\eta = 1\times 10^{-3}$, $\zeta = 2.85$ & 0 (0) & 0.15 (10) & 0.22 (10) \\
            bc & $1.4 \leq n \leq 1.9$, $\eta = 1\times 10^{-3}$, $\zeta = 2.85$ & 0 (0) & 0.17 (11) & 0.33 (15) \\
            \hline \hline 
        \end{tabular}
\caption{Similar to the slices in parameter space given in
     Tables~\ref{tab:Kappa_VS_ISM} and \ref{tab:Injec_VS_ISM},
     we show for the combination of the six choices of $g(n)$
     and $h(\eta)$ and the three choices for braking index
     $\kappa = 2.5, 3.0, 3.5$, the fraction of pulsar population
     simulations that are consistent within 3$\sigma$ with the
     \textit{AMS-02} positron-fraction spectrum. As before, in
     the parentheses the actual number of our simulations
     allowed are given.}
\vspace{-0.1cm}
\label{tab:Injec_VS_Kappa}
\end{table*}
 
We note that more data are being collected by the
\textit{AMS-02} experiment that can confirm these first
indications of preferences in the parameter space of pulsar population 
properties.  In particular, we expect that if a clear cut-off in the
positron fraction is observed above 500 GeV in energy, then a
significant fraction of the 205 pulsar astrophysical
realizations will be excluded.

\subsection{Including data from \textit{CALET} or \textit{DAMPE}}

Recently, two more satellite experiments, \textit{CALET}
\cite{2015JPhCS.632a2023A} and \textit{DAMPE}
\cite{TheDAMPE:2017dtc}, have published their measurements of
the total $e^{+}+e^{-}$ CR flux up to 5 TeV
\cite{Adriani:2018ktz, Ambrosi:2017wek}.  These spectra allow
us to test pulsar-population models at energies where their
expected fractional contribution to the total measured 
quantities can be very significant. For instance at 50 GeV, our
fits suggest that the pulsar population is responsible for
$\simeq 50 \%$ of the measured positron fraction value 
but only for $\simeq 8\%$ of the observed $e^{+}+e^{-}$ flux. At
500 GeV, pulsars can instead be responsible for as much as $ 90
\%$ of the positron fraction and up to $40\%$ of the
$e^{+}+e^{-}$ flux, with the pulsar contribution becoming
potentially even  more important at higher energies. 

\subsubsection{The highest energies: understanding the young,
nearby pulsars}
\label{sub:YoungCloseByPulsars}

\begin{figure*}[htbp]
\centering
\includegraphics[width=3.4in,angle=0]{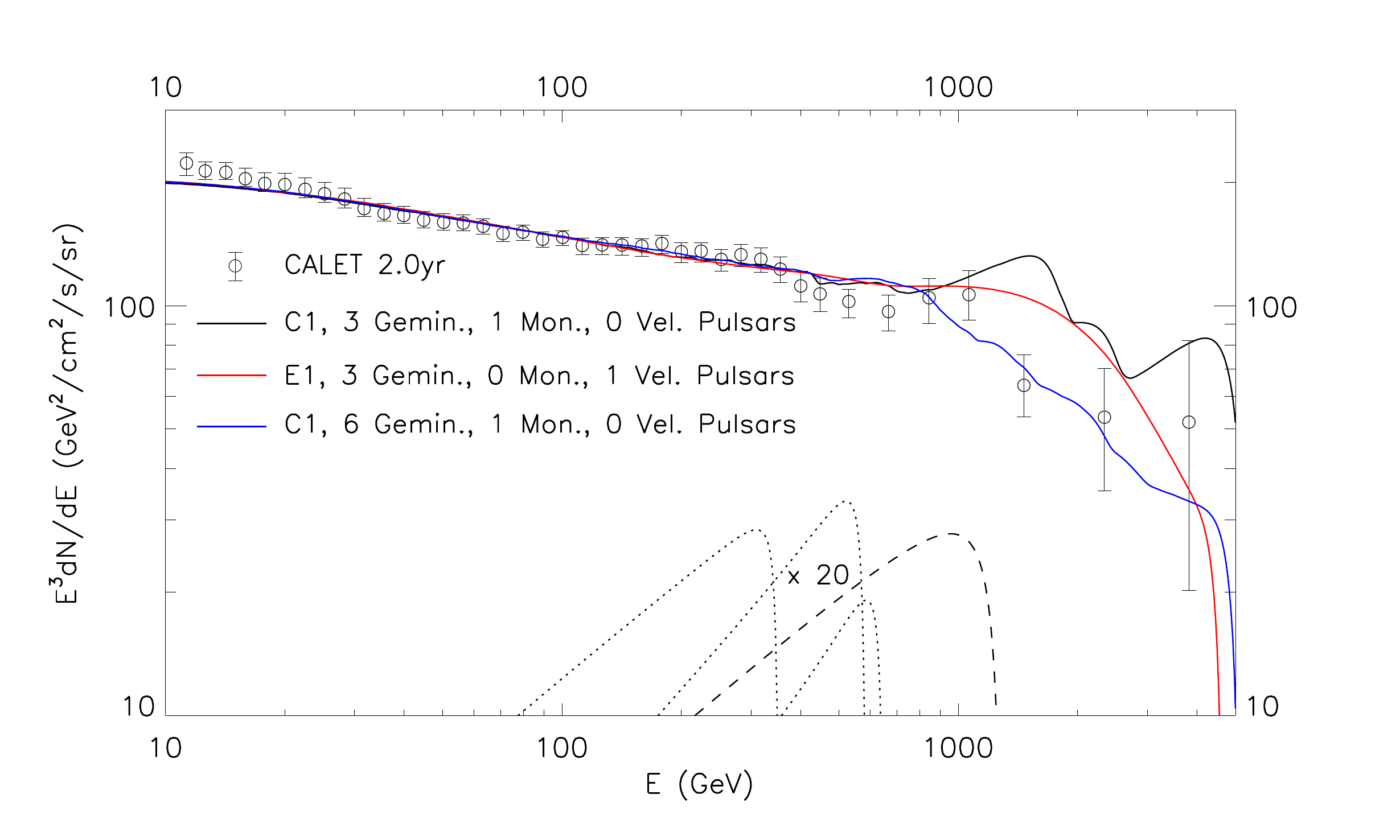} 
\includegraphics[width=3.4in,angle=0]{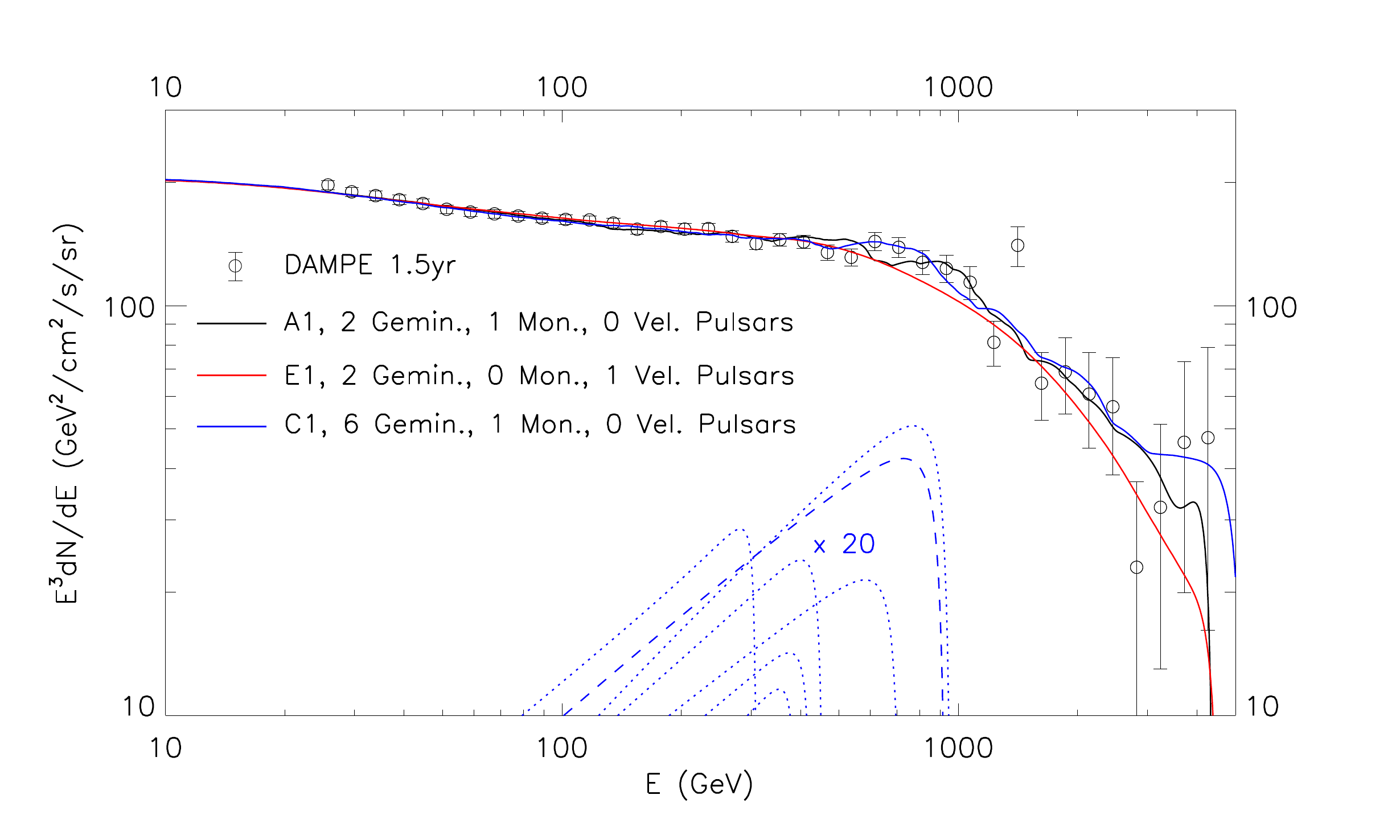}
\vspace{-0.4cm}
\caption{The simulated $e^{+}+e^{-}$ fluxes from combinations of
     pulsars. We include the primary $e^{-}$ and secondary
     $e^{\pm}$ CR components. On the \textit{left} panel we
     present three simulations that fit well the \textit{CALET}
     spectral data depicted, while in the \textit{right} panel
     we do the same with the \textit{DAMPE} spectral data.  For
     the simulations presented we give the number of pulsars
     that have age, distance from Earth, and current spin-down
     power in agreement with the observed ones of Geminga
     ("Gemin"), Monogem ("Mon") and Vela ("Vel") (see text for
     details). At high energies, we note the presence of several
     features coming from the contribution of individual pulsars
     and also a cut-off/change of slope to the total $e^{\pm}$ 
     flux above a TeV due to a small number of contributing sources. 
	 We also present for
     one simulation the fluxes from the individual pulsars that
     have  properties similar to those of Geminga
     (\textit{dotted} lines) and Monogem (\textit{dashed}
     lines). These fluxes are multiplied by a factor of 20 to
     lie within the panels' range. There are large
     source-to-source variations in our simulations even among
     Geminga-like pulsars, as we describe in the text.} 
\label{fig:CAL_DAM_Fluxes}
\end{figure*}

At higher energies, the combination of volume and age necessary for 
pulsars to be able to contribute is reduced. 
Thus the number of sources drops, with the 
highest energies probing only a small number of pulsars. This should 
result in $e^{+}+e^{-}$ fluxes rich in features as is shown in 
Fig.~\ref{fig:CAL_DAM_Fluxes}, where we compare some of our simulations to the 
data from \textit{CALET} (left) and \textit{DAMPE} (right). 
Additionally at $\simeq 1$ TeV, we find in our simulations that
even after smoothing  the possible small-scale features, there is
typically either a cut-off in the $e^{+}+e^{-}$ flux spectrum or
a change in its slope. This is attributed  to the fact that only
$O(10)$ pulsars can contribute at these energies. The exact
energy that this change in the spectrum occurs at and the
resulting spectral characteristics depend  on these pulsars'
individual properties. 
  
In order to understand the impact of some known nearby
pulsars, we compare with similar pulsars in our simulations that fall in
our neighborhood. There is extended literature on the possible
contribution from Geminga (J0633+1746) and Monogem (B0656+14) to
the high-energy positron flux \cite{Profumo:2008ms,  
Cholis:2013psa, Linden:2013mqa, Yin:2013vaa, Linden:2014sea, Hooper:2017gtd, 
Abeysekara:2017old, Yuan:2017ysv, Fang:2017tvj}. We test the
possible contribution from these pulsars by identifying
simulations that have pulsars with the same combination 
of distance from the Earth, age, and current spin-down
luminosity as these two pulsars.  More specifically, we check
for pulsars with (distance, age,  $\dot{E} (t = \textrm{age})$)  
that are within the (distance central value $\pm 1\sigma$, spin-down 
age$_{-50\%}^{+100\%}$, spin-down luminosity$_{-50\%}^{+100\%}$)
of the reported (0.25$_{-0.09}^{+0.45}$ kpc,  3.42$\times
10^{5}$ yr, 3.2$\times 10^{34}$erg/s) for Geminga and (0.29
$\pm$0.15 kpc, 1.11$\times 10^{5}$ yr, 3.8$\times 10^{34}$erg/s)
for Monogem \cite{Manchester:2004bp, ATNFSite}.  We note that the spin-down 
age and luminosity rely on the measurements of the period $P$ and its time derivative
$\dot{P}$ and are calculated assuming a braking index of $\kappa
= 3$.  Since we want to be agnostic about the true braking
index, we allow for the wide range around the central values of
age and luminosity. Furthermore, we check for simulations with
pulsars that have similar properties as Vela (B0833-45) (0.28
$\pm$ 0.14 kpc, 1.13$\times 10^{4}$ yr, 6.9$\times 10^{36}$erg/s). 

In both panels of Fig.~\ref{fig:CAL_DAM_Fluxes} we show the CR
spectra from three simulations that can explain the data. For
each of these simulations the number of pulsars that have
distance, age, and luminosity properties similar to Geminga, Monogem,
and  Vela are provided. It is typically easier to get pulsars
like Geminga and Monogem, while a pulsar like Vela is relatively
rare. Out of our simulations that fit the positron-fraction
data, 75$\%$ have at least one Geminga-like and 18$\%$ at least
one Monogem-like pulsar, while only 3$\%$ have at least one
Vela-like pulsar.

In each of the Fig.~\ref{fig:CAL_DAM_Fluxes} panels, for a
chosen simulation, we also plot the predicted $e^{+}+e^{-}$
fluxes from these particular pulsars. Since Vela-like pulsars
are younger, their contribution can be dominant only above a few
TeV, but the suppression of the total flux suggests that
Vela is not a dominant source of $e^{\pm}$ either because these
CRs have not yet reached us or because its $\eta$ is
suppressed. 
In fact recent work from Ref.~\cite{Evoli2018}  suggests that a 
significant fraction of high energy CRs from very young pulsars 
(as is Vela) are strongly confined for $O(10)$ kyr before being 
released  into the ISM. Monogem-like pulsars contribute at $\simeq 1$ TeV
energies and  Geminga-like pulsars at $\simeq 500$ GeV. The
exact cut-off energy depends on the ISM energy-losses rate and
the individual pulsar's age. Even after fixing the ISM model and
comparing only among the Geminga-like pulsars, there are
source-to-source variations in the cut-off due to the age
uncertainty that we include for these  sources. The amplitude
instead depends on their distance from Earth (that in turn has
variations since we allow an uncertainty on this as well) and
also on the $\eta$ parameter that is unique to each
source. Finally the slope for energies lower than the cut-off 
depends on the diffusion assumptions and on the
individual/unique pulsar injection index $n$ (for further
details see Ref.~\cite{Malyshev:2009tw}).

We note, that we do not over-plot the \textit{CALET} and
\textit{DAMPE} spectral data and we do not try to fit our
simulations simultaneously to both. The measured spectral indexes
of the $e^{+}+e^{-}$ flux are in some disagreement between these  
data-sets, suggesting some energy-scale-related measurement uncertainty. 

\subsubsection{Combining \textit{AMS-02} and \textit{CALET} data} 
\label{sub:CALET}

Instead, we combine the \textit{AMS-02} separately with the
\textit{CALET} data. We do that by taking all the 
original 900 simulations with their best $\chi^{2}$ fit
free-parameter values to the positron fraction and test their
predicted $e^{+}+e^{-}$ flux to the measurements 
by \textit{CALET} and \textit{DAMPE} respectively. We have
noticed that, as is also discussed in Ref.~\cite{Adriani:2018ktz},
the \textit{CALET} fluxes agree pretty well with the
\textit{AMS-02} data\footnote{We also tested the earlier
results by the \textit{CALET} collaboration published in
\cite{Adriani:2017efm}. These data also agree well with the
\textit{AMS-02} measurements.}. For each simulation, in fitting
to  the \textit{CALET} (\textit{DAMPE}) $e^{+}+e^{-}$ data, we
allow for an additional $10(20) \%$ freedom in the primary
$e^{-}$, secondary $e^{\pm}$ and pulsars $e^{\pm}$
normalizations starting from the best fit values to the positron
fraction\footnote{In theory we could have fitted the three
normalizations simultaneously to the \textit{AMS-02} $\&$
\textit{CALET} or \textit{AMS-02} $\&$ \textit{DAMPE}
spectra. We choose that approach instead, since from the
comparison of \textit{CALET} and \textit{DAMPE}, we understand
that energy-scale systematic errors between different
experiments can be important. We also allow for greater freedom
when combining with the \textit{DAMPE} data to reduce the
possible impact of these systematics. The 20$\%$  is also close
to the upper limit of some of these parameters
uncertainties.}. Since  all the data points are above 10 GeV in
energy, the impact of solar modulation is insignificant, and we do
not refit the relevant $\phi_{0}, \phi_{1}$ parameters.  

\begin{figure}
\hspace{0.35cm}
\includegraphics[width=3.4in,angle=0]{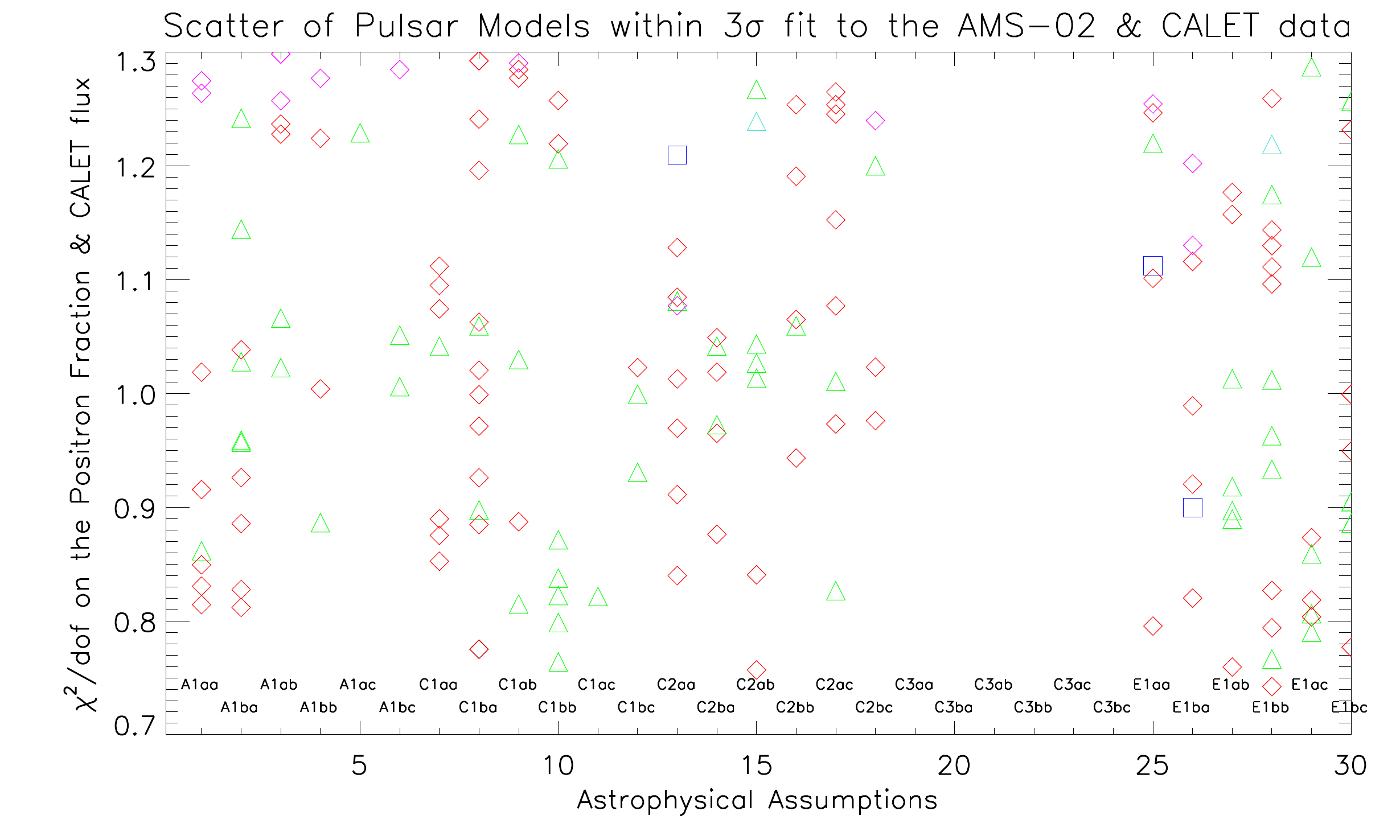}\\
\hspace{0.35cm}
\includegraphics[width=3.4in,angle=0]{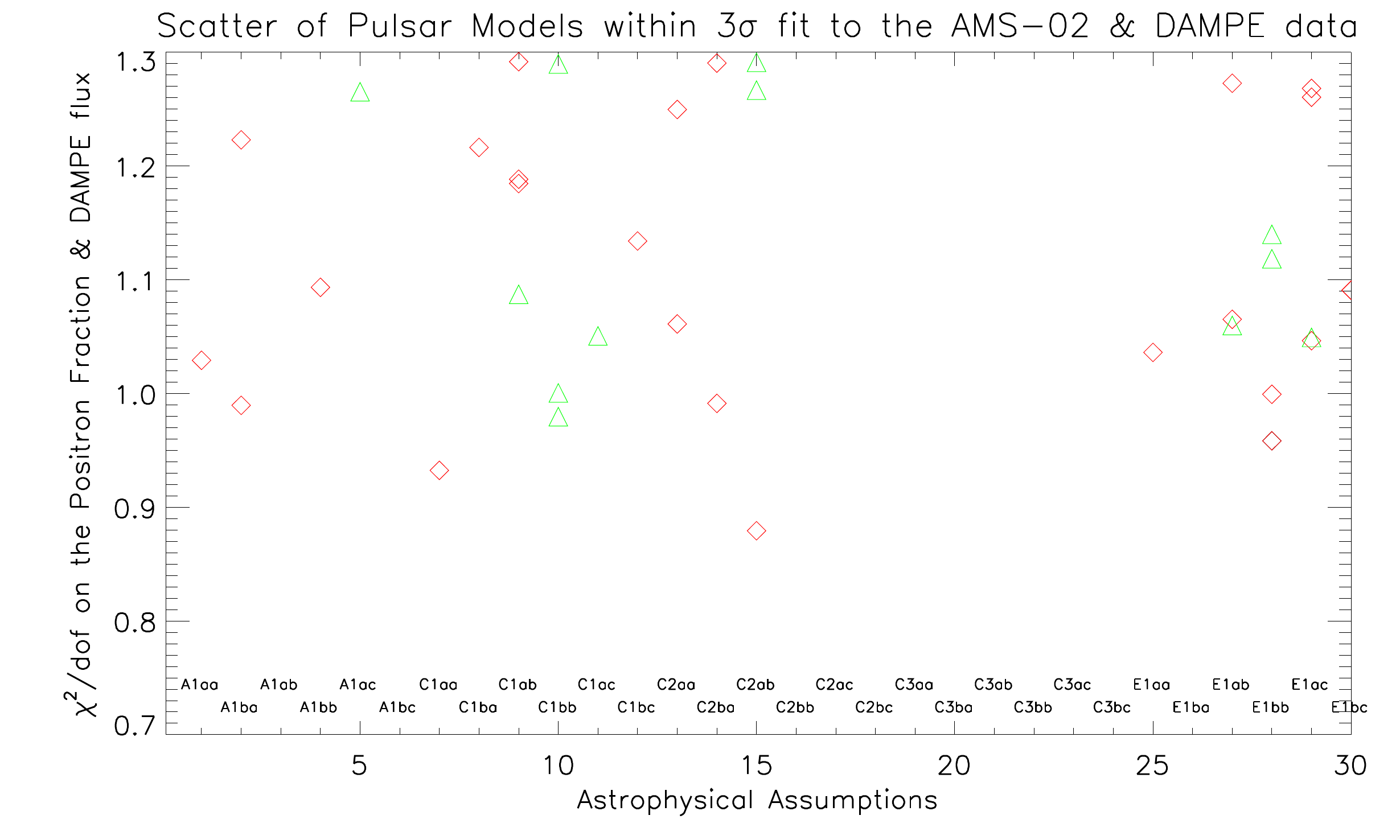} \\
\vspace{-0.2cm}
\caption{As in Fig.~\ref{fig:AMS2014scatter}, we present the
     scatter of astrophysical pulsar realizations that fall
     within our 3$\sigma$ constraint to the combined
     \textit{AMS-02} positron fraction and \textit{CALET}
     (\textit{top} panel) or \textit{DAMPE} (\textit{bottom}
     panel) $e^{+} + e^{-}$ flux data. The y-axis is the
     $\chi^{2}$/d.o.f.\ value for the combined data (see text for
     details), while the x-axis describes the  same
     astrophysical assumptions as in
     Fig.~\ref{fig:AMS2014scatter}. For each of the three values
     of braking index that we test, we depict realizations by
     two colors depending on the quality of their fit to the
     positron fraction alone. For braking index $\kappa$ of 3.0,
     the red diamonds are for simulations allowed within
     3$\sigma$ by the positron-fraction data. These same
     simulations are among the 205 simulations described in
     Section~\ref{sec:AMSfit}. With magenta diamonds we
     depict models allowed within 3$\sigma$  by the combined
     data, but that are excluded by \textit{AMS-02} data
     alone. Similarly, for $\kappa = 3.5$, green (turquoise)
     triangles show the realizations allowed (excluded) by the
     \textit{AMS-02} positron fraction. No additional models
     with $\kappa = 2.5$ are allowed by adding the
     \textit{CALET} measurement. With \textit{DAMPE} data there
     are no blue, magenta and turquoise symbols.} 
\label{fig:AllLeptonsScatter}
\end{figure}

In Fig.~\ref{fig:AllLeptonsScatter} (top panel), we give the
scatter of pulsar realizations that are in agreement with the
\textit{AMS-02} and \textit{CALET} data within 3$\sigma$ from
what is a $\chi^{2}$/d.o.f. of 1\footnote{There are 51
\textit{AMS-02} data points fitted by 5 degrees of freedom and
39 \textit{CALET}  or \textit{DAMPE} data points fitted by 3
additional parameters. For the remaining 82 degrees of freedom for 3$\sigma$, the combined $\chi^{2}$/d.o.f =
1.288.}.  The axes are the same as in
Fig.~\ref{fig:AMS2014scatter} with the exception that the y-axis
now refers to the combined data fit. As in
Section~\ref{sec:AMSfit} we portray the three tested values for
the braking index $\kappa$  with there different symbols (boxes,
diamonds and triangles for the simulations with  $\kappa = 2.5$,
3.0 and 3.5). Comparing to the $e^{+}+e^{-}$ flux data, it is easy
to acquire a good fit. In fact, of the 205 simulations that are
allowed within the 3$\sigma$ threshold by the positron fraction
data, 145 are also allowed by the combined positron fraction and
$e^{+}+e^{-}$ flux data. As in
Fig.~\ref{fig:AMS2014scatter}, those are shown by the blue boxes,
red diamonds, and green triangles, respectively. Some  
pulsar astrophysical realizations have a good enough fit to
the \textit{CALET} data and although not allowed within the
3$\sigma$ threshold by only \textit{AMS-02}, are allowed by the
combined data. There are 14 such simulations in addition to the
145. We show these by magenta diamonds and turquoise triangles
for $\kappa = 3.0$ and 3.5 respectively, which typically lie on
the top end of the y-axis. No additional models with $\kappa =
2.5$ were added in the analysis by including the \textit{CALET}
data.

We find that, as with the positron fraction fit, ISM model "C3"
is systematically excluded by the data regardless of the other
astrophysical assumptions. In fact, the addition of the flux data 
does not alter the conclusions of Section~\ref{sec:AMSfit} or set 
further preferences among the remaining ISM models. Similarly, it
is still the case that only a small number of simulations with
$\kappa = 2.5$ (3 out of 159) are allowed, and beyond that,
adding the flux data only slightly affects the preference for
simulations with $\kappa = 3.0$ versus 3.5. Regarding the
fraction of the spin-down power going into CR $e^{\pm}$, the
preference for narrower $h(\eta)$ is sustained, while the $g(n)$
index of injection distribution properties conclusions are
unaffected by the \textit{CALET} data.

\begin{figure}
\vspace{-0.5cm}
\hspace{0.35cm}
\includegraphics[width=3.4in,angle=0]{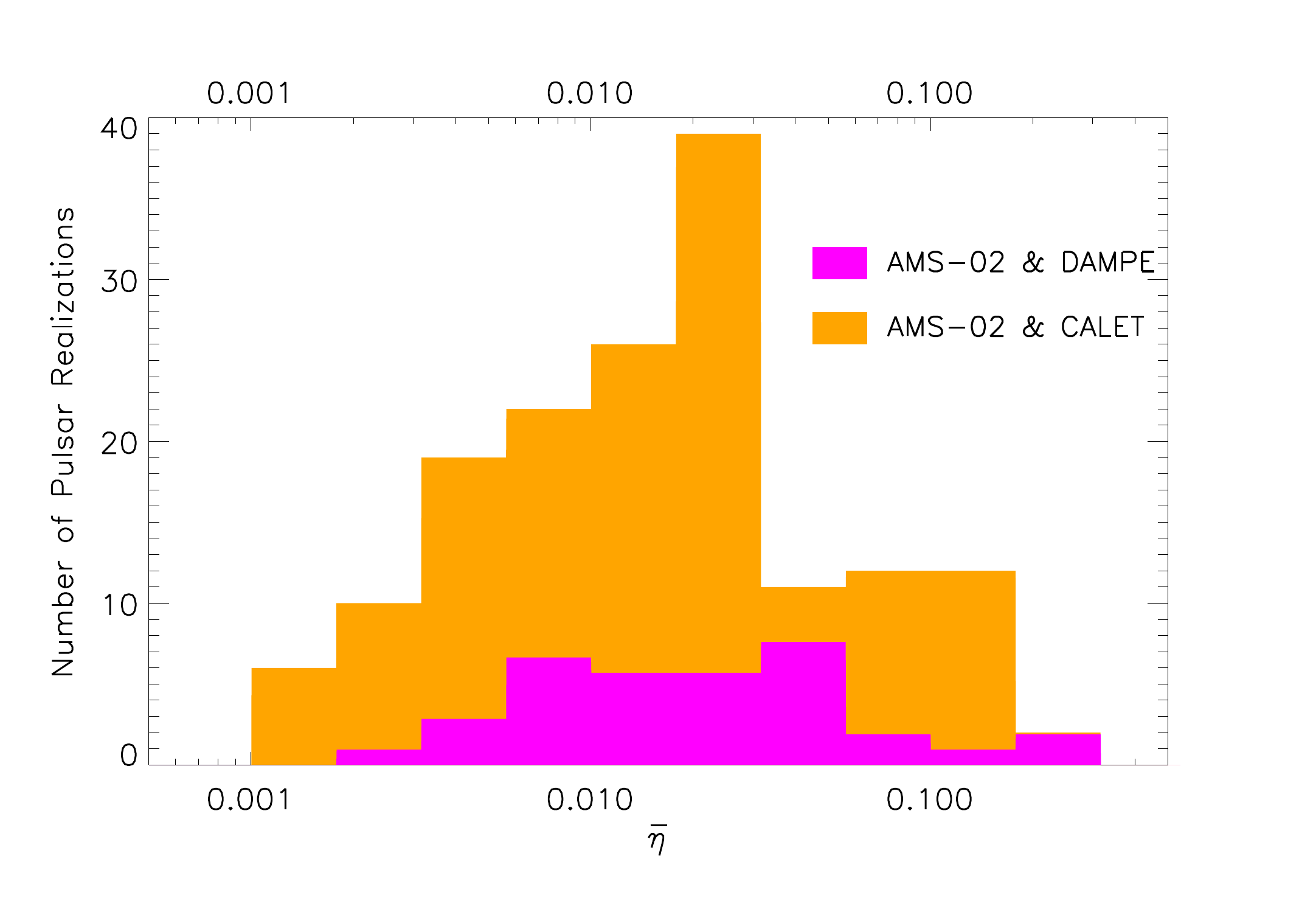}
\vspace{-0.6cm}
\caption{The distribution of the fitted values of the
    $\bar{\eta}$ parameter for the pulsar-population
    simulations, allowed within 3$\sigma$ by the combination of
    \textit{AMS-02} $\&$ \textit{CALET}  data (159 astrophysical
    pulsar realizations in the orange histogram), or allowed by
    the \textit{AMS-02} $\&$ \textit{DAMPE} data (36 in the
    magenta histogram).}
\label{fig:Eta}
\end{figure}

In Figure~\ref{fig:Eta}, we also present a histogram of the
fitted values of the mean fraction $\bar{\eta}$ of the pulsars
population from each realization. The number on the y-axis is
the number of these pulsar astrophysical realizations that are
allowed.  The range of $\bar{\eta}$ to explain the CR data is
between $10^{-3}$ and 0.26, with a clear peak at 1--3$\times
10^{-2}$. These fractions assume a Milky Way pulsar birth 
 rate of one per century. If that rate is doubled, those fractions
are smaller by the same factor.

\subsubsection{Combining \textit{AMS-02} and \textit{DAMPE} data}

\label{sub:DAMPE}
On the other hand, combining the \textit{DAMPE} $e^{+} + e^{-}$ flux data
with the \textit{AMS-02} positron fraction dramatically cuts
down the parameter space of the pulsar astrophysical
realizations allowed (see bottom panel of
Fig.~\ref{fig:AllLeptonsScatter}).  Out of the 205 simulations
allowed by the positron-fraction data, only 36 remain after  
combining with \textit{DAMPE}, while the down-weighting of the
\textit{AMS-02} data does not lead to any additional simulations
allowed within the 3$\sigma$ threshold.  We find that none of
the $\kappa = 2.5$ simulations survive, while 2/3 of these 
astrophysical realizations are with $\kappa = 3$. Regarding the
impact of these data on the ISM and $h(\eta)$, $g(n)$
distributions: our results discussed above hold given the
small number of remaining simulations. The range of $\bar{\eta}$
is basically the same as that from $\textit{CALET}$ and
\textit{AMS-02} data (Fig.~\ref{fig:Eta}). 

We finally note that tracking the source of the systematic
difference between \textit{CALET} and \textit{DAMPE} $e^{+} +
e^{-}$ data is of great importance in properly combining the
data in future analyses. 

\section{Discussion and Conclusions}
\label{sec:conclusion}

We have used CR $e^{\pm}$ data to study the properties of the
population of local Milky Way pulsars. Using the \textit{AMS-02}
positron-fraction measurement and the $\textit{CALET}$ and
\textit{DAMPE} $e^{+} + e^{-}$ fluxes, we have tested 
hundreds of pulsars simulations.  These simulations probe the
astrophysical uncertainties associated with the CR leptonic spectra
observed at Earth.  In particular, we
produce 900 simulations that sample a broad range of
possibilities for the spatial distribution of neutron-star birth
locations in the Milky Way, for their ages, and their spin-down
properties. We also study the effects of uncertainties in the
spectra of CR $e^{\pm}$ injected into the ISM and those
associated with CR propagation through the ISM.  We describe how
we model all these astrophysical effects and how we combine them
in Section~\ref{sec:method}.
 
We work under the assumption that pulsars are the dominant
source for the rise of the positron fraction at high
energies. We find that consistent pulsar-population models can
result in a continued rise in the positron fraction with energy,
a flattening at energies beyond 300 GeV, or even a drop-off.  We
also find that consistent models can produce a positron-fraction
spectrum that is either smooth in energy or that has
fluctuations with energy (see Fig.~\ref{fig:FourPulsarModels}).

We find ISM models with energy losses that are
suppressed relative to what is conventionally assumed
\cite{Moskalenko:2001ya, GALPROPSite,DRAGONweb},  
but that are still allowed within reasonable uncertainties on
the local magnetic and interstellar-radiation fields, are
excluded by either the \textit{AMS-02} positron fraction data
alone, or in combination with \textit{CALET} or
\textit{DAMPE} data.  The underlying reason for this result is
that low energy losses cause the pulsar spectra (even after
performing a $\chi^{2}$ fit) to overshoot the current spectral
data at high energies (where the energy losses have a dominant
impact). Simulations probing other typical ranges of ISM
conditions are also performed as described in
Section~\ref{sec:method}, with the relevant results in
Section~\ref{sec:results}.
 
Furthermore, a pulsar braking index $\kappa$ of 2.5 or less is
disfavored by the data, regardless of most of the other
astrophysical assumptions, with a very small fraction of our
simulations with $\kappa = 2.5$ being allowed. In this work we
consider  pulsars to have a constant braking index throughout
their time evolution, and set their population spin-down
properties so that we can explain the radio-frequency
observations from many such sources. These radio-pulsars are
typically a Myr old. A  braking index of 2.5 produces pulsars
that are very powerful sources at their younger stages, when most of
the CR $e^{\pm}$ are produced. Consequently, high energy data that 
probes younger pulsar sources is overshot by these simulations. 
We also find some small preference for a braking index of 3
versus 3.5, from the CR data. These  results are also given in
Figs.~\ref{fig:AMS2014scatter} and~\ref{fig:AllLeptonsScatter}
and Tables~\ref{tab:Kappa_VS_ISM} and~\ref{tab:Injec_VS_Kappa}.

Given that we have observed several very young pulsars (with
ages of $O(10^{4})$ yrs or less) with a braking index of less
than 3, these results create a tension. One solution to that, 
would be that pulsars have time-variable braking indexes. 
For instance, pulsars might start with smaller indexes
leading to fast spin-down in their initial stages. As the
braking index increases to a value of $\simeq$3, relevant for
magnetic-dipole radiation, the pulsars spin down more slowly. If
the braking index indeed changes with time, we will need
additional simulations to probe all the possible paths of
$\kappa (t)$, versus the $\kappa = \textrm{constant}$ studied in
this work. 

Since pulsars are observed to have large variations in their
surface magnetic fields and original periods, we
study the resulting distribution of their initial spin-down
power as a population.  We find no indications for any
preference in terms of a narrow or a wide initial spin-down
power distribution.  Also, the fraction of rotational energy
going to CR $e^{\pm}$ is very uncertain. By modeling it with a
log-normal distribution, we find that the CR data fits hint at
a narrow distribution for this fraction (see
Figs.~\ref{fig:AMS2014scatter} and~\ref{fig:AllLeptonsScatter}
and Tables~\ref{tab:Injec_VS_ISM}
and~\ref{tab:Injec_VS_Kappa}). In terms of the averaged (of the
pulsar-population) value, the fraction is fitted in half of our
simulations to be $\simeq 1 -3 \%$ (see Fig.~\ref{fig:Eta}). The
exact injection spectral properties of pulsars are still not
well constrained by the CR data.

The CR spectra above $\simeq 500$ GeV, accessible currently by
\textit{CALET} and \textit{DAMPE}, allow us to study young nearby
pulsars. We find that above a TeV in energy, pulsar simulations
can explain the observed change in the $e^{+}+e^{-}$ slope,
since the number of contributing sources drops to only
$O(10)$. However, as this is a small number of pulsars, large
variations in the predicted CR spectra are seen between
simulations, associated with the properties of the individual 
contributing pulsars.  Finally, the $e^{\pm}$ fluxes that are 
from young, nearby, and energetic pulsars (like Vela) are 
constrained by the data.
 

\begin{acknowledgments}  We thank Joseph Gelfand, Victoria
Kaspi, Ely Kovetz, Tim Linden, Dmitry Malyshev and Robert Schaefer for 
valuable discussions. This research project was conducted using computational
resources  at the Maryland Advanced Research Computing Center
(MARCC) and supported by NASA Grant No.\ NNX17AK38G, NSF Grant
No.\ 0244990, and the Simons Foundation.
\end{acknowledgments}  

\bibliography{Pulsars_Properties}

\end{document}